\documentclass[%
 aip,
 amsmath,amssymb,
 reprint,%
]{revtex4-1}

\usepackage{graphicx}
\usepackage{dcolumn}
\usepackage{bm}

\usepackage[utf8]{inputenc}
\usepackage[T1]{fontenc}
\usepackage{mathptmx}

\newcommand{\Zn}{\text{Z}}

\newcommand{\Hn}{\text{H}}
\newcommand{\gn}{\text{g}}
\newcommand{\mx}{\text{max}}

\newcommand{\vv}{\text{v}}

\newcommand{\ef}{\text{eff}}

\usepackage{url}
\usepackage[version=3]{mhchem}
\usepackage{color}
\usepackage{amsmath}
\usepackage{lipsum}
\usepackage{amssymb}
\usepackage[version=3]{mhchem}
\usepackage{mathrsfs}      
\usepackage{booktabs}   
\usepackage{subfigure}
\usepackage{float}
\usepackage{appendix}
\usepackage{comment}

\usepackage{epsfig}
\usepackage{dcolumn}%
\usepackage{relsize}

\begin{document}

\title{Consistent Kinetic-Continuum Dissociation Model II: Continuum Formulation and Verification} 

\author{Narendra Singh}
 \email{singh455@umn.edu.}
\author{Thomas Schwartzentruber}%
\affiliation{ 
Department of Aerospace Engineering and Mechanics, 
University of Minnesota, Minneapolis, MN 55455}%


\date{\today}

\begin{abstract}In this article, we implement  a recently developed non-equilibrium chemical kinetics model \cite{singhmodeldevelopment2019} based on \textit{ab initio} simulation data and perform verification studies. Direct molecular simulation data is used to verify the predictive capabilities of the model. Using the model, dominant physics such as need for a rotational energy equation, and the quantitative role of non-Boltzmann effects are identified. Based on the analysis and reasonable assumptions, a simplified model for implementation into large-scale CFD simulations is proposed. Without incurring  additional computational cost, the model can be used in existing flow solvers to analyze hypersonic flows.
\end{abstract}

\maketitle

\section{Introduction}
Non-linearities and multi-scale physics make the modeling of hypersonics flows a complex problem. The flow first experiences a strong shock which partially dissociates the gas. Partially dissociated gas then flows around the vehicle, enters the boundary layer, and induces gas surface chemistry. In numerical predictions, uncertainty in the thermochemical state leads to uncertainty in other processes such as heat flux to the surface, gas surface reactions etc. State-of-the-art models used in numerical simulations are based on limited experimental data \cite{byron1966shock,appleton1968shock,hanson1972shock} with large uncertainities. For instance, in computational fluid dynamic simulations (CFD), the widely used Park model \cite{park1989assessment} has known limitations and is not consistent with recently obtained ab-initio results \cite{valentini2016dynamics,bender2015improved,grover2019jtht,chaudhry2019statistical,panesi2014pre,macdonald2018construction_DMS, andrienko2018vibrational}. In order to reduce uncertainty, ab-initio calculations using computational chemistry methods have resulted in massive amounts of data. Quasi-classical trajectory (QCT) calculations and direct molecular simulations (DMS) have created huge databases of rate constants simulating billions of collisions for air species. 

The purpose of the current research is to capture the ab-initio trends with the simplest model that still retains accuracy. In a related article \cite{SinghCFDI} a new model was proposed that captures the state-specific cross-sections with simple functions that can be analytically integrated to obtain a continuum model.
The model embeds key physics such as dependence on vibrational energy, rotational energy, relative translational energy, centrifugal barrier, bound and quasi-bound effects. 
The model systematically incorporates non-Boltzmann effects \cite{Singhpnasplus}, the details of which can be found in Ref.~\cite{SinghCFDI}. In this article, we verify the new model vs Direct Molecular Simulation (DMS) calculations. Since, the new model is analytically derived from kinetic level rates and therefore has different terms containing contributions from different physics, we study the relative importance of each term towards overall dissociation.  For example, is an additional evolution equation for rotational energy necessary to incorporate its effects on reaction rates? Quantitatively, what are the effects of non-Boltzmann distributions? More specifically, what is the contribution of over-population of the high energy states in the early phase of excitation and the depletion of high internal energy states in the quasi-steady state (QSS) phase? What is the relative influence of diatom-atom interactions to diatom-diatom interactions? Finally, based on the study we make recommendations for model simplifications that retain accuracy and  can be easily incorporated into state-of-the-art CFD.

\section{Continuum Framework for Internal Energy Transfer and Reaction Chemistry} \label{Results_and_Discussion}

In order to verify the accuracy of the new dissociation model, and in order to study the relative importance of various nonequilibrium mechanisms, we perform zero-dimensional thermo-chemical relaxation calculations. Such calculations essentially involve the source terms in the energy transport equations and species continuity equations solved by state-of-the-art hypersonic CFD codes. Such zero-dimensional calculations are therefore relevant for large scale CFD codes and, at the same time, are useful to verify accuracy compared to Direct Molecular Simulation (DMS) calculations. 
Specifically, we consider a nitrogen gas characterized by translational temperature ($T$), rotational temperature ($T_{rot}$) and vibrational temperature ($T_{vib}$). $T_{rot}$ and $T_{vib}$ are ``psuedo''-temperatures such that by prescribing Boltzmann distributions of internal energy corresponding to these temperatures, one recovers the average rotational ($\langle \epsilon_{rot} \rangle$) and average vibrational ($\langle \epsilon_{v} \rangle$) energy of the gas. The system of zero-dimensional equations include source terms for species concentrations, rotational energy, vibrational energy and total energy:
  \begin{equation}
    \frac{ d [N_2]}{dt} = -k_{N_2-N_2} [N_2][N_2] - k_{N_2-N} [N_2][N]
     \label{Rate_eqn_pop}
 \end{equation}
   \begin{equation}
    \frac{ d [N]}{dt} = -2\frac{ d [N_2]}{dt} 
     \label{Rate_eqn_pop1}
 \end{equation}
   \begin{equation}
   \begin{split}
    \frac{ d \langle \epsilon_{rot} \rangle}{dt} = \frac{\langle \epsilon_{rot} ^* \rangle-\langle  \epsilon_{rot}  \rangle}{\tau_{\text{mix,rot}}} -k_{N_2-N_2} [N_2] (\langle \epsilon_{rot}^d \rangle - \langle \epsilon_{rot}  \rangle) \\ - k_{N_2-N} [N] (\langle \epsilon_{rot}^d \rangle  - \langle \epsilon_{rot}  \rangle) 
    \end{split}
     \label{Jeans_modified}
 \end{equation}
 
 \begin{equation}
  \begin{split}
    \frac{ d \langle \epsilon_v \rangle}{dt} = \frac{\langle \epsilon_v^* \rangle-\langle  \epsilon_v \rangle}{\tau_{\text{mix,v}}} -k_{N_2-N_2} [N_2] (\langle \epsilon_{v}^d \rangle - \langle \epsilon_v \rangle) \\- k_{N_2-N} [N] (\langle \epsilon_{v}^d \rangle - \langle \epsilon_v \rangle) 
    \end{split}
     \label{LandauTeller_modified}
 \end{equation}

\begin{equation}
   E = \cfrac{3}{2} k_B T \left( [N_2] +[N] \right)+ [N_2] (\langle  \epsilon_v \rangle+\langle  \epsilon_{rot} \rangle) +[N] \cfrac{\epsilon_d}{2}
   \label{total_energy}
\end{equation}

In the above equations, $[N_2]$ and $[N]$ denote the concentration of molecules and atoms in the gas, and $k_{N_2-X}$ is the overall dissociation rate constant with collision partner $X$ ($N_2$ or $N$). Furthermore $\langle \epsilon_v^*\rangle$  and $\langle \epsilon^*_{rot} \rangle$ are the average equilibrium (corresponding to T) vibrational and rotational energies respectively, and $\langle \epsilon_v^d \rangle$ is the average vibrational energy of dissociating molecules, and $\langle \epsilon_{rot}^d \rangle $ is the average rotational energy of dissociating molecules. E is the total energy density of the molecules used to obtain $T$ in adiabatic relaxation. In the Landua-Teller and Jeans equation terms, $\tau_{\text{mix,v}}$ is the mixture vibrational relaxation time constant and $\tau_{\text{mix,rot}}$ is the mixture rotational relaxation time constant given by \cite{Lee1984}:
\begin{equation}
    \tau_{mix,i} = \cfrac{[N2] +[N]}{ \cfrac{[N2]}{\tau_{N_2-N_2,i}}+\cfrac{[N]}{\tau_{N_2-N, i}}}
\end{equation}
 where $\tau_{N_2-N_2, i}$ is relaxation time constant due to collision of N$_2$ and N$_2$, $\tau_{N_2-N, i}$ is relaxation time constant due to collision of N$_2$ and N and $i$ refers to either rotation or vibration.
 
The \textit{key model parameters} required by the equations, and therefore required by state-of-the-art thermochemical nonequilibrium CFD codes, are the relaxation time constants ($\tau_{\text{mix,v}}$ and $\tau_{\text{mix,rot}}$), the dissociation rate coefficient expression ($k_{N_2-X}$), and the average internal energy lost due to dissociation ($\langle \epsilon_v^d \rangle $ and $\langle \epsilon_{rot}^d \rangle$).
 
\begin{figure}
  \subfigure[Isothermal relaxation using master equation and DMS in Ref.~\cite{erikrobynnitro}]{
   \includegraphics[width=3.0in,trim={0.10cm 0.1cm 0.0cm 0.1cm},clip]{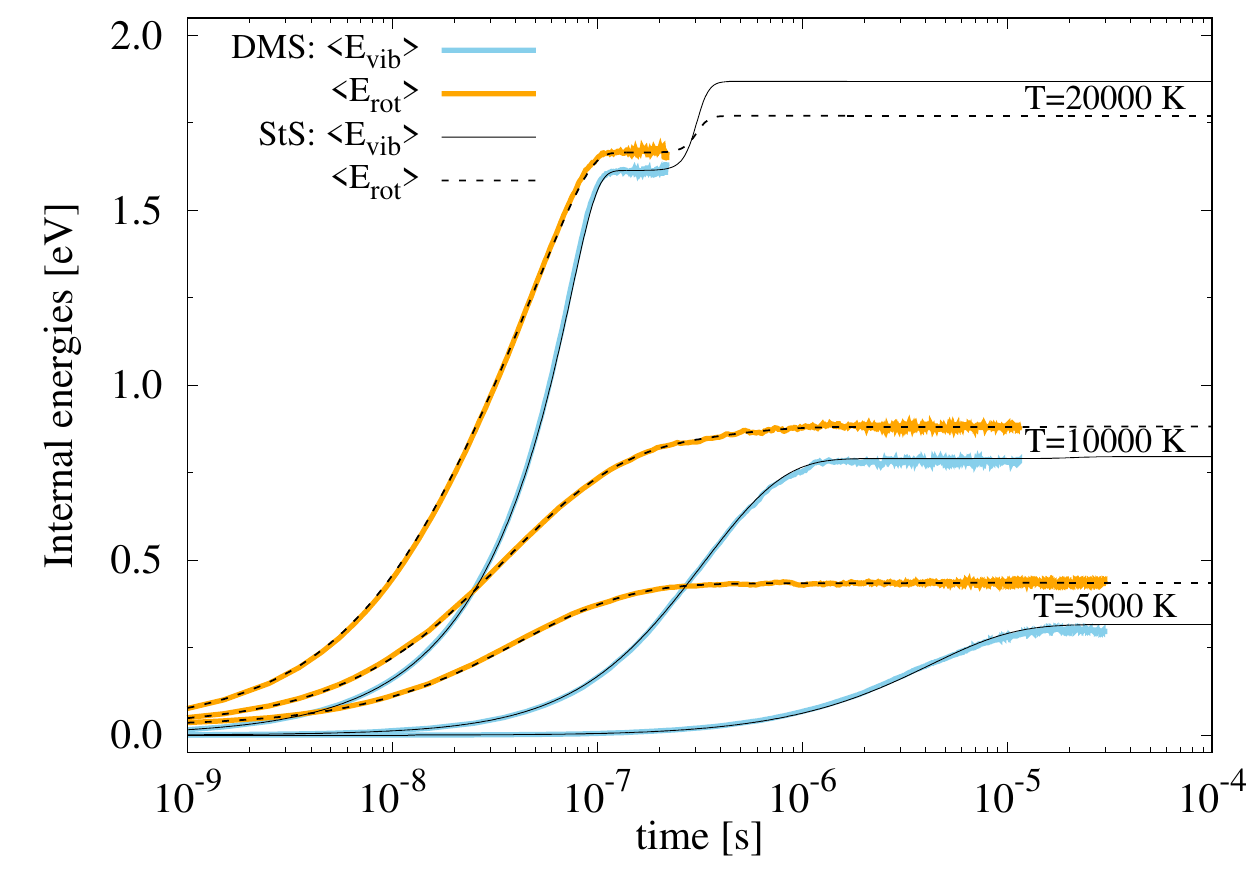}
    \label{comp_intro}
  }
\subfigure[Vibrational relaxation time constants in Ref.~\cite{valentini2016dynamics}]{
\includegraphics[width=3.2in,trim={0.10cm 0.1cm 0.1cm 0.1cm},clip]{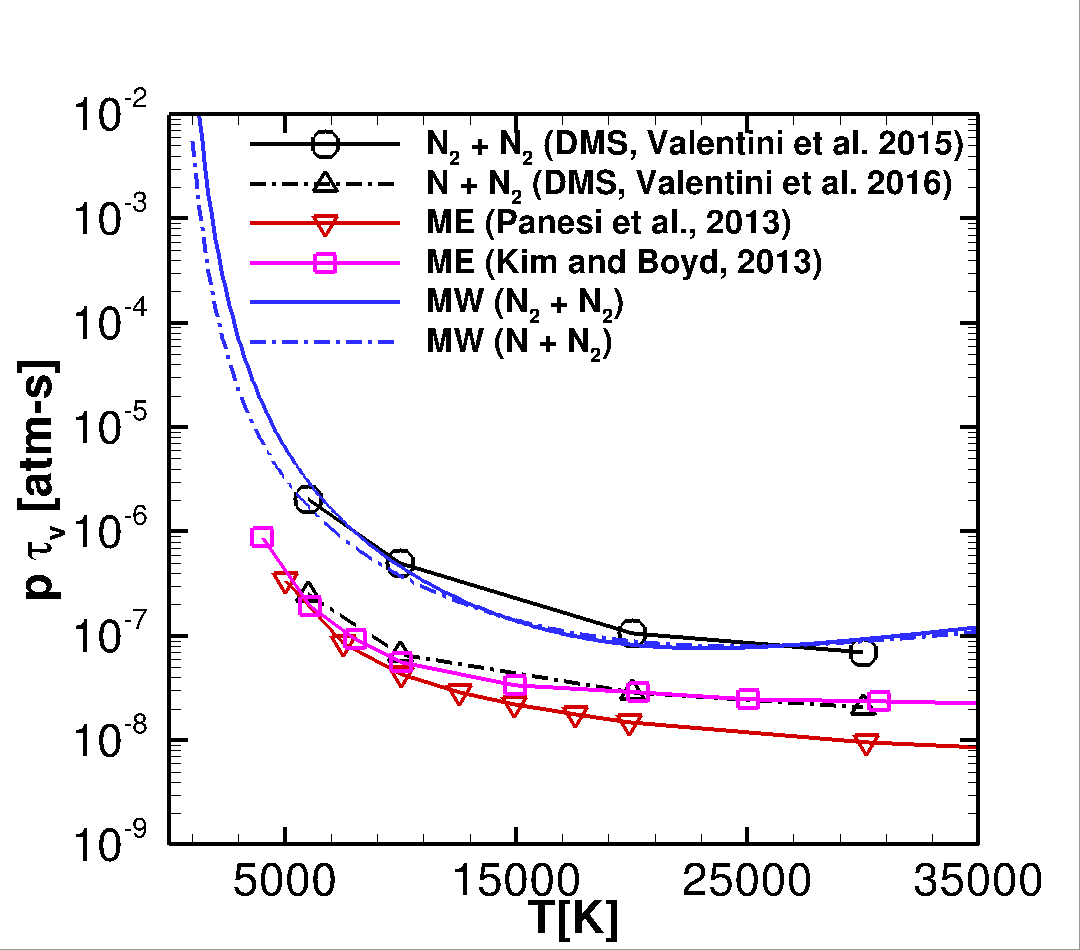}
 \label{time_constant_vib}
}
 \caption{ 
 Isothermal ro-vibrational excitation and dissociation in a reactor for N$_2$-N interactions via DMS and master equation (StS) analysis  (b) Vibrational relaxation time constant via DMS \cite{valentini2016dynamics,valentini2015N4}, state-resolved master equation analysis \cite{panesi2014pre,kim2013state} and Millikan-White experimental data \cite{millikan1963systematics}.}
  \label{DMS_intro}
\end{figure} 

Time constants for rotational relaxation and vibrational relaxation have been computed by master-equation studies for N+N$_2$ collisions \cite{panesi2014pre,kim2013state}, and by DMS calculations for \emph{both} N+N$_2$ and N$_2$+N$_2$ collisions \cite{valentini2016dynamics}. Figure~\ref{comp_intro} shows master-equation and DMS calculations of rovibrational excitation under isothermal conditions as reported in Ref. \cite{erikrobynnitro}. As seen in Fig.~\ref{comp_intro}, the DMS and full state-t-state (StS) results are identical, leading to the confidence in the N-N$_2$ relaxation rates. Relaxation time constants for translational-rotational and translational-vibrational energy transfer can be inferred from a range of isothermal simulations corresponding to different translational temperature (such as those in Fig.~\ref{comp_intro}). As an example, Fig. \ref{time_constant_vib} shows the temperature variation in $\tau_v$ for both N+N$_2$ and N$_2$+N$_2$ collisions, as first reported in Ref. \cite{valentini2015direct}.  Also included in Fig. \ref{time_constant_vib} are the correlations from Millikan and White (MW)\cite{millikan1963systematics} that are widely used in CFD codes. The MW correlations are based on experimental data for N$_2$+N$_2$ collisions and the MW correlation for N+N$_2$ collisions was assumed to be very similar. In contrast, DMS and master-equation calculations clearly show that $\tau_v$ is an order-of-magnitude lower (faster relaxation) for N+N$_2$ collisions. In this article, new model results will use the $\tau_v$ and $\tau_{rot}$ values inferred from DMS calculations. These new model results will be compared to the state-of-the-art CFD model, which consists of the MW correlations for $\tau_v$ and the Park $T$-$T_v$ dissociation model. Details of state-of-art CFD model are given in the appendix (Ref.~\ref{Existing_CFD}).

It is important to note that the Landau-Teller expression is mathematically derived under certain strict assumptions, such as mono-quantum transitions, and does not explicitly model multi-quantum transitions that occur frequently in exchange collisions (i.e. N$^a$N$^b$ + N$^c \rightarrow$ N$^a$N$^c$ + N$^b$). This could be investigated more closely at the kinetic scale, using DSMC for example. However, as demonstrated in this article, we find that as long as the time-constant expressions are fit to the new ab-intio data (DMS or master-equation results), that the Landau-Teller expression is able to accurately capture the evolution of average internal energy. 

 
The new dissociation rate model, $k(T, \langle \epsilon_{rot} \rangle, \langle \epsilon_v \rangle )$, was derived in a separate article by the authors \textcolor{red}{\cite{singhmodeldevelopment2019}}. The model consists of a compact expression that accurately captures the state-resolved dissociation cross-sections computed by QCT calculations (i.e. probabilities of dissociation given the translational, rotational, and vibrational energies involved in a collision). This expression is then integrated over a simple model for non-Boltzmann distribution functions \cite{Singhpnasplus} of internal energy to provide an analytical, closed-form, dissociation rate expression for use in multi-temperature CFD codes. Since the new continuum dissociation rate expression is analytically derived from the kinetic model equations, it includes a number of rather complicated terms representing various nonequilibrium mechanisms. Such mechanisms (physics) include the coupling between average translational, rotational, and vibrational energy with dissociation, anharmonic effects in the diatomic potential energy surface, separate contributions due to bound and quasi-bound molecules, as well as non-Boltzmann internal energy populations such as overpopulation and depletion of high-energy tails during different stages of the gas evolution. The full expression for $k(T, \langle \epsilon_{rot} \rangle, \langle \epsilon_v \rangle )$ is listed in the Appendix, along with the values of all model parameters. 

The new model for the average internal energy lost due to dissociation ($\langle \epsilon_v^d \rangle$ and $\langle \epsilon_{rot}^d \rangle$) was also derived analytically from the same cross-section information in Ref. \textcolor{red}{\cite{singhmodeldevelopment2019}}. This expression also includes the same nonequilibrium mechanisms (physics) as included in the dissociation rate expression, and the full equation for $\langle \epsilon^d_v \rangle$ is listed in the Appendix. While a similar analytical expression for $\langle \epsilon^d_{rot} \rangle$ could be used, we find that the simple expression $\langle \epsilon^d_{rot} \rangle = \epsilon_d - \langle \epsilon^d_{v} \rangle$ is accurate, and is also consistent with the findings of Bender et. al. \cite{bender2015improved}. This model for $\langle \epsilon^d_{rot} \rangle$ is also listed in Eq.~\ref{avg_edrot_gen} in sec.~\ref{avgedvrotappend} of the appendix.

Using the new model expressions for the three main parameters in Eqs.\ref{Rate_eqn_pop}--\ref{total_energy} , $\tau_{\text{mix,v}}$ and $\tau_{\text{mix,rot}}$, $k(T, \langle \epsilon_{rot} \rangle, \langle \epsilon_v \rangle )$, $\langle \epsilon_v^d \rangle$ and $\langle \epsilon_{rot}^d \rangle$, the purpose of this article is to verify the accuracy of the new models and to quantify the relative importance of different nonequilibrium mechanisms towards the overall thermochemical evolution of the gas. Specifically, section 3 compares predictions from the new model with baseline ab-intio results using the DMS method and also compares with predictions using the widely used Park and MW models. In Section 4, contributions to dissociation due to non-Boltzmann distributions of rotational and vibrational energy are quantified along with contributions from the average vibrational energy due to differences between $\tau_v$ for N+N$_2$ and N$_2$+N$_2$ collisions. After the relative contributions of various mechanisms (various terms) is quantified in Section 4, recommendations for model simplifications are proposed in Section 5, and the resulting simplified model form is compared to previous models from the literature. A preliminary model to include recombination is presented in Section 6. Conclusions of the research are summarized in Section 7.

\section{Comparison of Model Results with Direct Molecular Simulation}
In order to assess the accuracy of the new models, we compare with baseline Direct Molecular Simulation (DMS) calculation results. The DMS calculations include both N-N$_2$ and N$_2$-N$_2$ collisions, where all possible energy transitions are allowed and the PES  \cite{paukku2013global,paukku2014erratum} is the sole model input. A full description of the DMS method is contained in Ref.~\cite{tomDMS}. Since the kinetic models for state-resolved cross-sections and non-Boltzmann internal energy distributions were formulated using such DMS and QCT results using the same PES \cite{SinghCFDI,Singhpnasplus}, it is relevant to compare predictions from the new continuum model to these baseline ab-intio results. 

\begin{figure}
\centering 
    \subfigure[]
 {
 \includegraphics[width=3.2in]{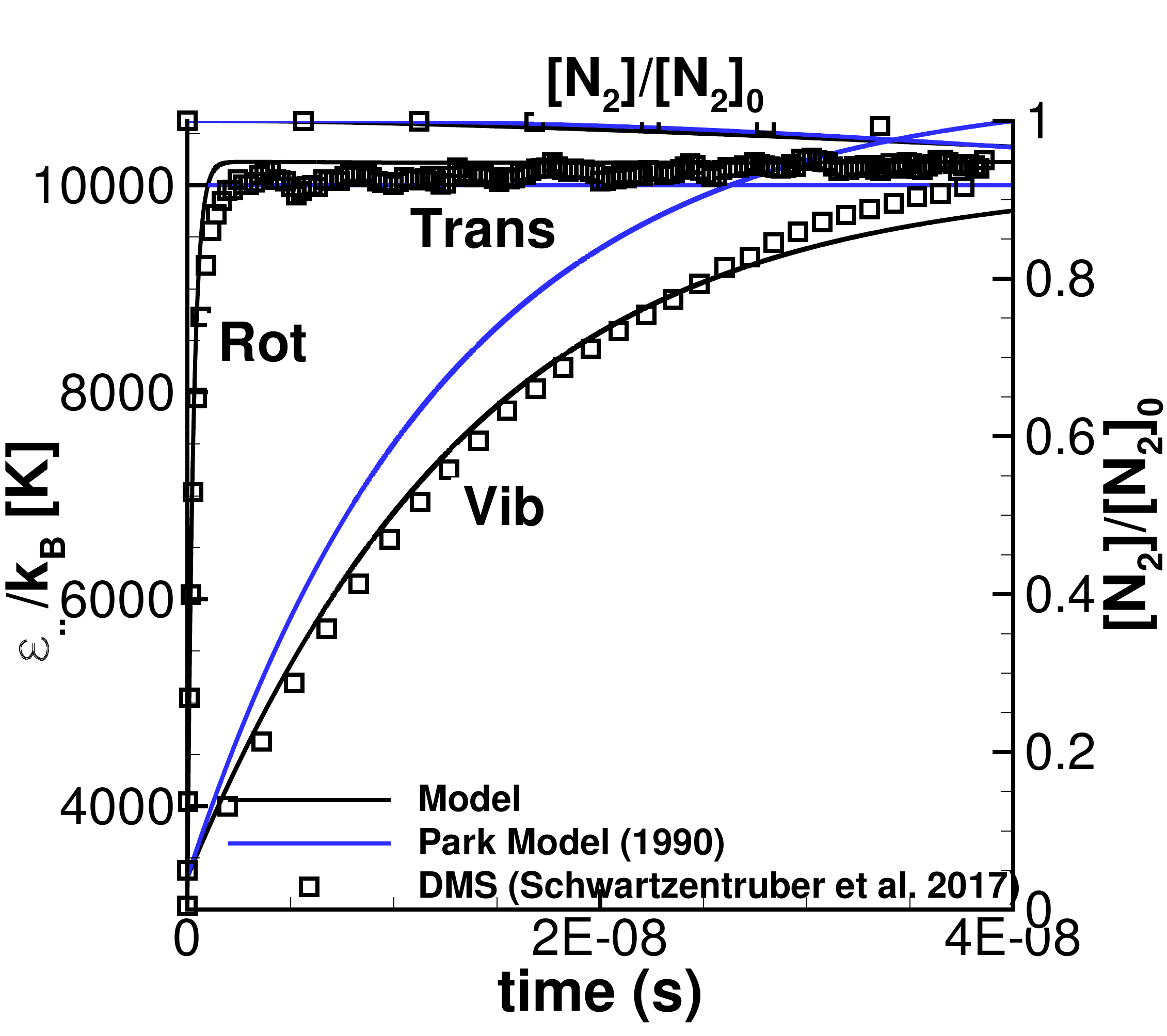}
 \label{10k-excite}
 }
    \subfigure[]
 {
     \includegraphics[width=3.2in]{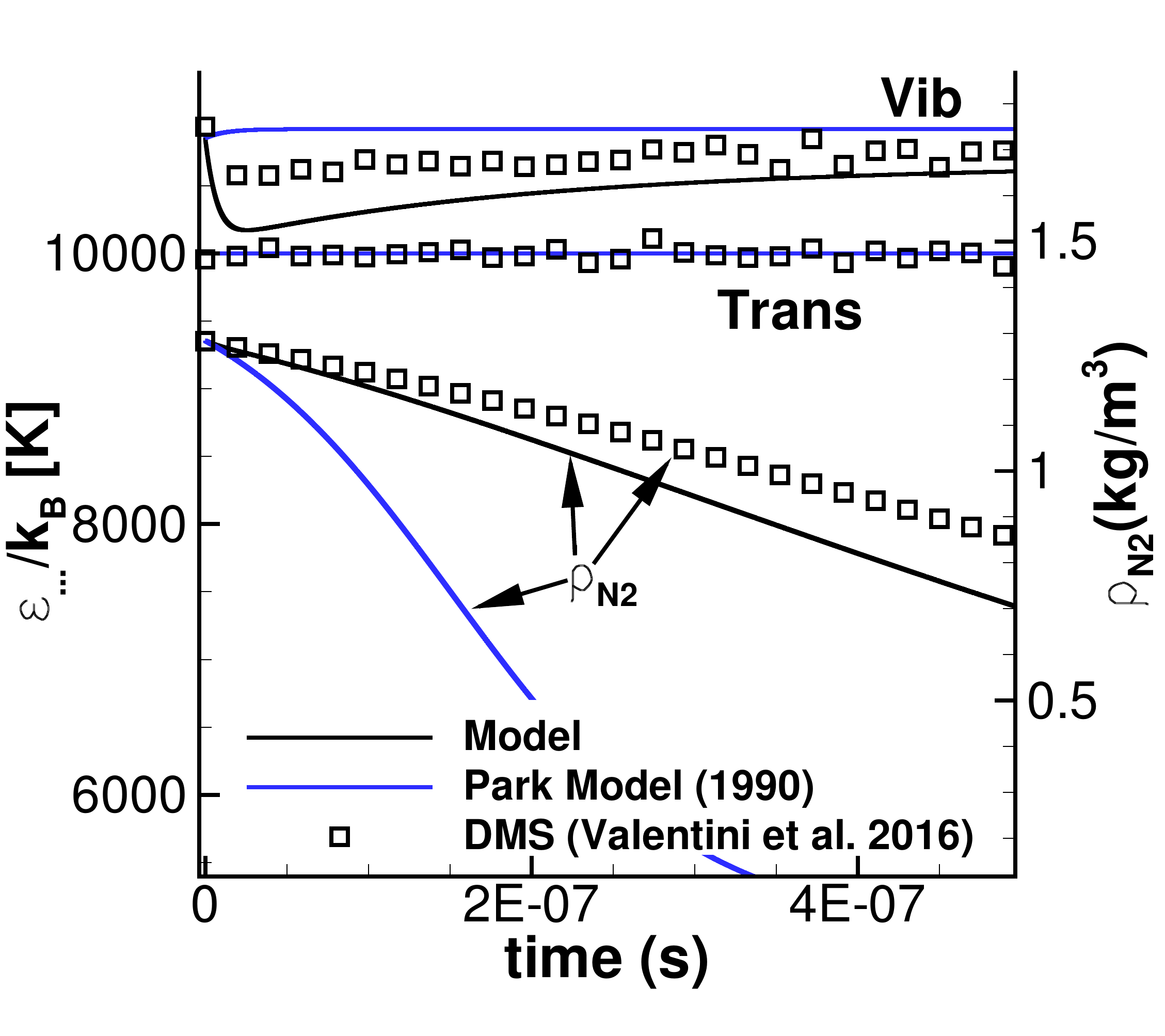}
   \label{10k-equil}
 }  
   \label{zerod_20K_Park}
   \caption{(a) Isothermal ro-vibrational relaxation of nitrogen gas at $T=10, 000$ K initialized from low ro-vibrational energy ($T_v=T_{rot} =2,000 K$) (b) Isothermal ro-vibrational relaxation of nitrogen gas at $T=10, 000$ K initialized at thermal equilibrium ($T_v=T_{rot} =10,000 K$). Results are obtained from DMS \cite{tomDMS}, Park model \cite{park1989assessment}, and the proposed model. } 
   \label{10k}
\end{figure}

The first set of comparisons are carried out under isothermal conditions. In the first case, the gas is initialized as N$_2$ molecules with high center-of-mass translational energy corresponding to $T$=10,000 K (kept constant during the calculation) and with internal energies corresponding to $T_{v}=T_{rot}=$2000 K. Full details of the DMS calculation for these conditions can be found in Ref. \cite{tomDMS}. Figure \ref{10k-excite} shows the DMS results where the rotational energy rapidly equilibrates with the translational energy, while the vibrational energy excites at a slower rate. During the vibrational excitation period, only 5\% of the gas has dissociated ($[N_2]/[N_2]_0$ at $t=4\times10^{-8}$s $\approx 0.95$). As seen in Fig. \ref{10k-excite}, the new model matches the DMS results closely. Additionally, results using the Park $TT_v$ model and MW vibrational relaxation time are shown in Fig. \ref{10k-excite}, where the rotational energy is assumed to be in equilibrium with the translational energy (the standard assumption in current CFD codes). For this case, while there is noticeable difference in the vibrational relaxation rate, the Park model results also match the DMS dissociation results closely. 

In the second case, the gas is initialized as N$_2$ molecules corresponding to $T$=10,000 K (kept constant during the calculation), but now with internal energies also initialized to high energy, corresponding to $T_{v} = T_{rot} = $10,000K. This shortens the calculation time for the DMS method to capture significant dissociation and also initiates the gas in translational-rotational equilibrium. Full details of the DMS calculation for these conditions can be found in Ref. \cite{valentini2016dynamics}. As seen in Fig. \ref{10k-equil}, compared to DMS, the new model predicts slightly faster dissociation and slightly more variation in the vibrational energy (rotational energy is not shown). The initial drop in average vibrational energy is caused by the fact that the gas is initialized using Boltzmann distributions for rotational and vibrational energy. As the gas immediately begins to dissociate, since molecules with high vibrational energy are exponentially more likely to dissociate, significant vibrational energy is removed from the gas. As shown in detail in Ref. \cite{valentini2016dynamics}, the gas then reaches a Quasi-Steady-State (QSS) where the vibrational energy distribution becomes time-invariant and includes depletion of the high energy states. This is the reason why in Fig. \ref{10k-equil} the vibrational energy trend initially decreases before reaching a plateau at longer times. Since the new model captures vibrationally favored dissociation including QSS depletion effects and the correct internal energy lost due to dissociation, the model agrees well with the DMS results. In contrast, as seen in Fig. \ref{10k-equil}, the Park $TT_v$ model predicts significantly faster dissociation and exhibits a constant vibrational energy trend. This is because the standard implementation of the Park model assumes the vibrational energy lost due to dissociation corresponds to the average vibrational energy of the gas (no preferential dissociation of high energy states is modeled). This results in a higher vibrational energy value leading to a higher dissociation rate (proportional to $\sqrt{TT_v}$) and the Park model also does not explicitly account for depletion effects. Finally, the Park model also predicts faster dissociation compared to the new model since the equilibrium dissociation rate (i.e. when $T=T_{rot}=T_v$) is slightly higher for the Park model than for the new ab-intio calculations \cite{valentini2016dynamics}. 

\begin{figure}[ht]
\centering 
    \subfigure[Isothermal Relaxation]
 {
 \includegraphics[width=3.2in]{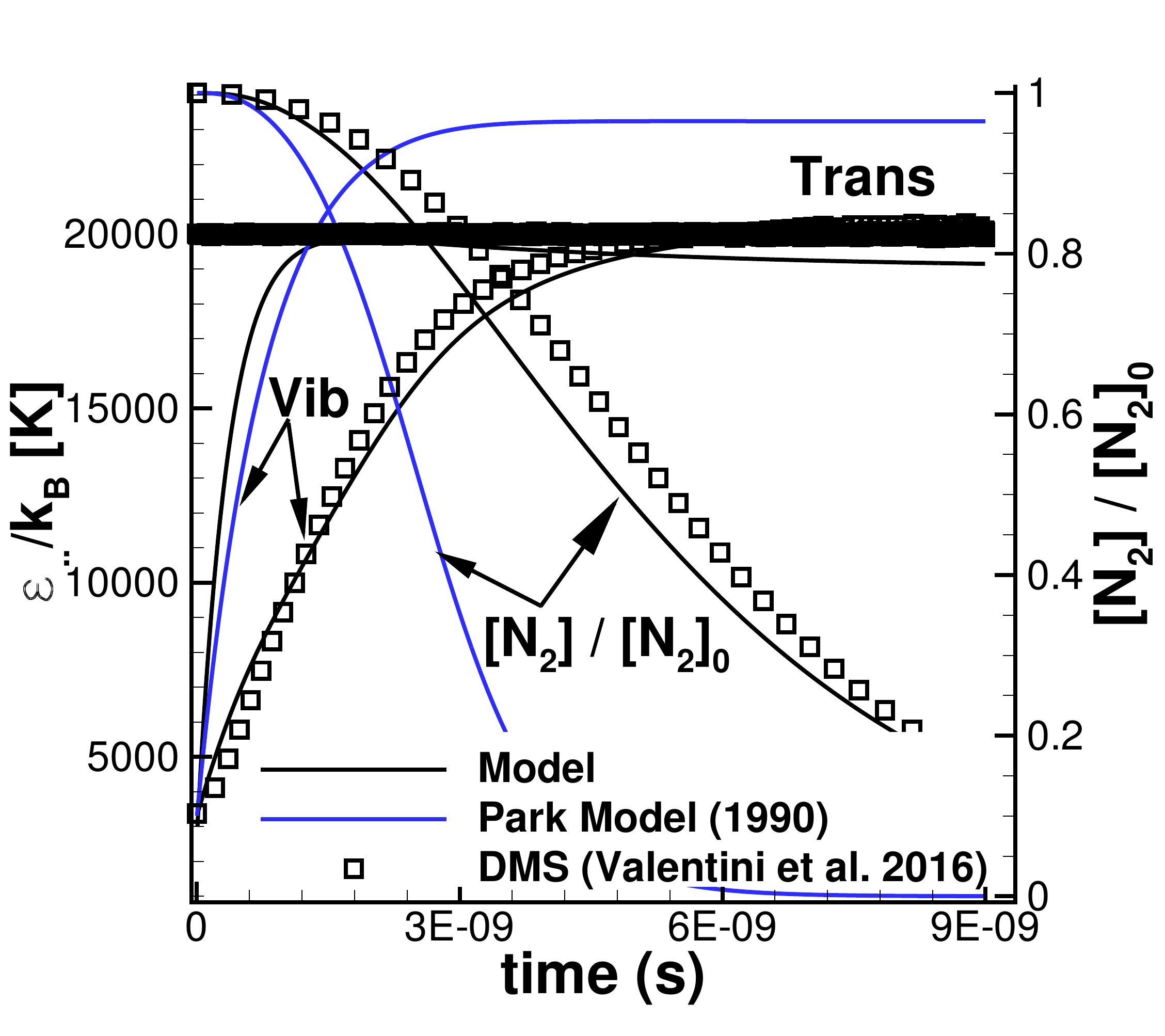}
 \label{20k-excite}
 }
    \subfigure[Adiabatic Relaxation]
 {
     \includegraphics[width=3.2in]{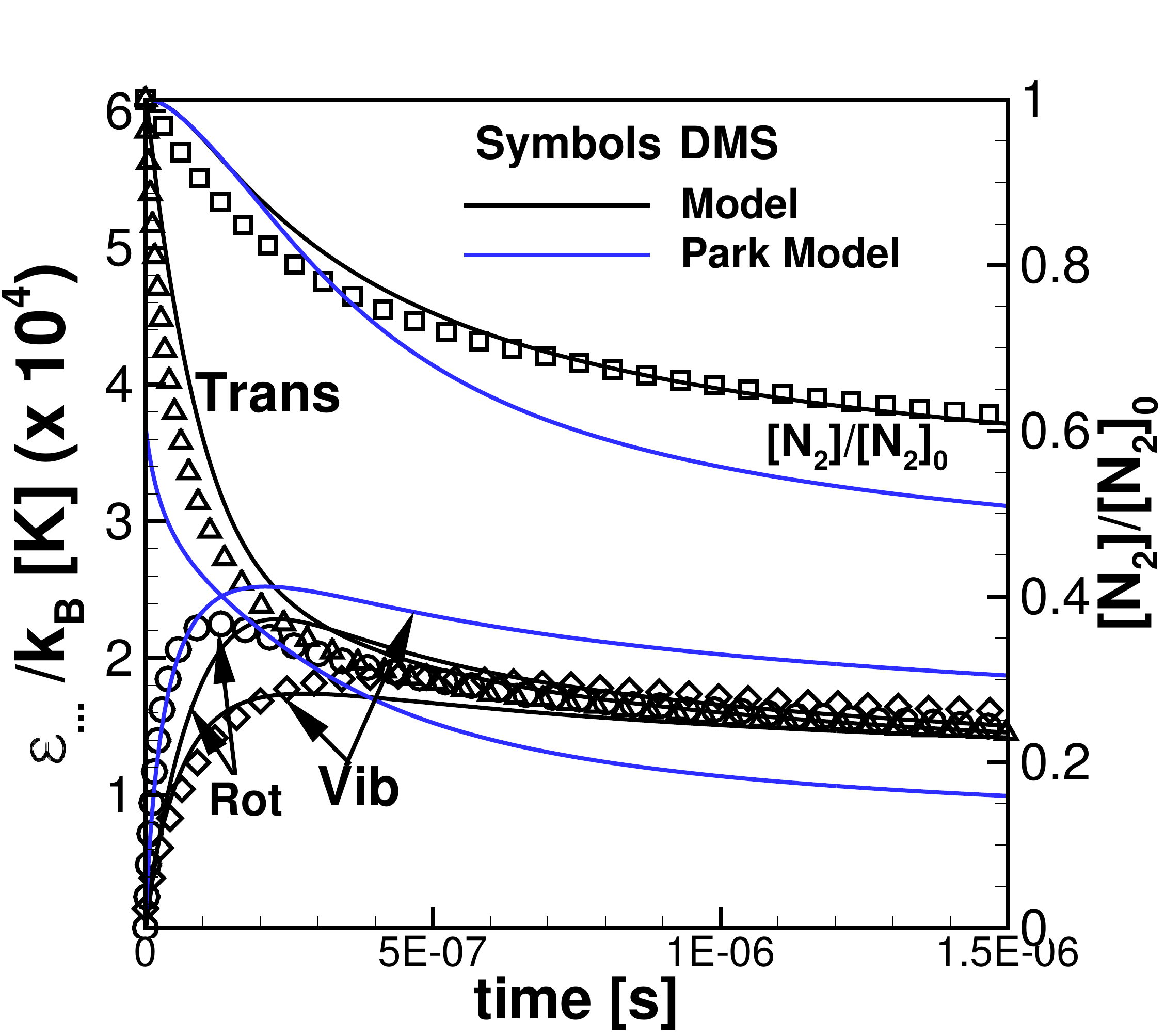}
   \label{15k-adiabatic}
 }  
   \label{20k}
   \caption{(a) Isothermal ro-vibrational relaxation of nitrogen at $T=20, 000$K. DMS results are taken from Ref.~\cite{valentini2016dynamics}. (b) Adiabatic relaxation of nitrogen gas. DMS results are taken from Ref.~\cite{torres2019direct}. Prediction from the proposed model and Park model \cite{park1989assessment} are also shown.} 
   \label{zerod_20K_N3_Full}
\end{figure}

In the third case, the gas is initialized as N$_2$ molecules with translational energy corresponding to $T$=20,000 K (kept constant during the calculation) and with internal energies corresponding to $T_{v}=T_{rot}=$3000K. Full details of the DMS calculation for these conditions can be found in Ref. \cite{valentini2016dynamics}. Figure \ref{20k-excite} shows the DMS results where significant dissociation occurs \emph{while} the vibrational energy is exciting towards the translational energy (for clarity, rotational energy is not shown). The new model predictions agree closely with the baseline DMS result. In contrast, the Park model using MW relaxation time constants (refer to Fig. \ref{time_constant_vib}) predicts a much higher average vibrational energy profile and significantly faster dissociation. There are several reasons for this disagreement. First, the initial vibrational relaxation mainly involves N$_2$-N$_2$ collisions, and the MW time-constant value is lower than the ab-initio value used in the new model (see Fig. \ref{time_constant_vib} at 20,000 K). Second, as discussed in the previous paragraph, as the gas dissociates the Park model does not remove sufficient vibrational energy from the gas (i.e. no preferential dissociation from high energy levels), which in-turn increases the dissociation rate. As a clarification, the reason the average vibrational energy equilibrates at a higher value than the average translational energy (for the Park result) is simply due to the relation between temperature and average energy \cite{SinghCFDI}. Specifically, if Fig. \ref{20k-excite} was plotted using pseudo temperatures, the Park model temperatures ($T$ and $T_v$) do equilibrate, while for the DMS and model results, $T_v < T$ in the QSS region due to the depletion of high-energy states. Similar trends in the average energies, compared to the trends in pseudo temperatures, are evident in many of the figures in this article.

In the fourth case, the gas is initialized as N$_2$ molecules with translational energy corresponding to $T$=60,000 K and internal energies corresponding to $T_{v}=T_{rot}=$300K, but now the simulation is performed under adiabatic (constant total energy) conditions. The static enthalpy ($30.1$ MJ/kg) involved in this case is representative of post-shock conditions at a flight velocity of 7.8 km/s. Full details of the DMS calculation for these conditions can be found in Ref. \cite{torres2019direct}. As seen in Fig. \ref{15k-adiabatic}, the DMS calculations (symbols) show the gas immediately starts to dissociate and translational energy is rapidly removed from the system. Rotational energy rapidly excites into equilibrium with the translational energy, while the vibrational energy takes longer to approach equilibrium, after which energy in all three modes is gradually reduced as dissociation continues. While the 20,000 K isothermal case shown in Fig. \ref{20k-excite} may seem extreme, Fig. \ref{15k-adiabatic} shows that immediately behind such a strong shock wave, a considerable amount of dissociation (approximately 30\%) occurs while the translational temperature is upwards of 20,000 K and internal energy is rapidly exciting. The remainder of the dissociation, under these conditions, occurs in the QSS phase (where $T \approx T_{rot} \approx T_v$) characterized by depleted internal energy distribution functions, as described in detail in Ref. \cite{torres2019direct}. As evident from Fig. \ref{15k-adiabatic}, the new model agrees closely with the baseline DMS results. Similar to the trends observed in the first three cases, the Park model does not remove sufficient vibrational energy due to dissociation and over-predicts the dissociation rate as a result. To clarify, since the Park model assumes trans-rotational equilibrium, the initial condition is set as $T_{tra-rot}= 36, 627$ K such that the overall energy in the trans-rotational mode is equivalent to that in the DMS calculation.

One interesting feature of the Park model is that the vibrational energy becomes frozen significantly above the translational energy (this is also evident if plotted as pseudo temperatures). This overshoot in vibrational temperature (energy) has been observed in many prior hypersonic CFD simulations. This overshoot occurs because only the Landau-Teller translational-vibrational relaxation source term is present in the Park model, while the term accounting for removal of vibrational energy due to dissociation is zero (since $\langle \epsilon_v^d \rangle = \langle \epsilon_v \rangle$). Under strong shock conditions, the drop in translational energy due to dissociation is too rapid for the vibrational temperature to equilibrate with. Furthermore, since the MW relaxation rate slows down as the translational temperature drops, the translational-vibrational energy exchange further slows resulting in a rather prolonged overshoot for this case. The DMS and new model predictions do not exhibit such a noticeable overshoot. 

\section{Relative Importance of Various Nonequilibrium Mechanisms}

In the section, we explore the relative contribution of each nonequilibrium mechanism towards the overall dissociation process using the new model. Since the new model is analytically derived from kinetic-level expressions, the full functional form (listed in the Appendix) is rather complicated compared to the Park model. However, one benefit of such an analytically consistent, kinetic-continuum, model is that it contains separate terms and parameters associated with each physical mechanism. This allows us to quantify the influence of various mechanisms by comparing or removing certain model terms. The purpose is to determine if simplifications to the model can be made while maintaining model accuracy.
For example, a separate rotational energy equation not typically included for hypersonic CFD simulations, despite research showing rotational energy is important for dissociation \cite{bender2015improved,macdonald2018_QCT}. We first explore the role of rotational energy coupling to dissociation and whether it is necessary to track rotational energy evolution separately in Section \ref{role_rotation}.  
In Section \ref{role_nb},  we quantify the effect on dissociation due to the depletion of high vibrational energy states during QSS and due to the over-population high energy states during rapid excitation. 
Finally, in Sec.~\ref{N3_N4}, we study the sensitivity of the dissociation process to the details of N-N$_2$ collisions versus N$_2$-N$_2$ collisions, which involves the influence of exchange reactions on the gas vibrational relaxation.

\subsection{Role of Rotational Energy}\label{role_rotation}
\begin{figure}
\centering 
  {
    \includegraphics[width=3.5 in]{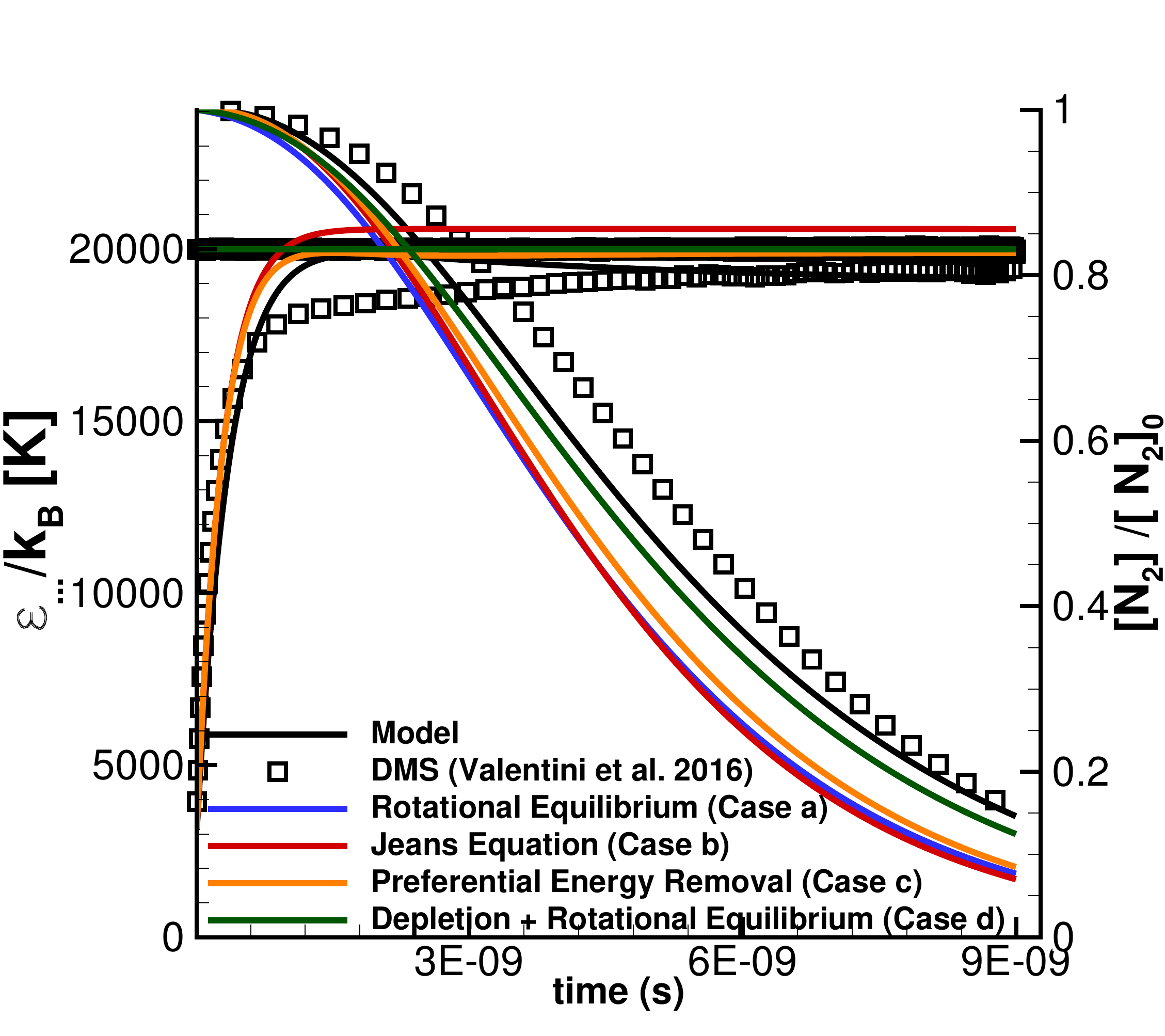}
   }  
   \caption{Isothermal relaxation of nitrogen at $T=20,000$ K to assess the role of rotational energy in dissociation. } 
   \label{ROT}
\end{figure}

Due to the fast relaxation of rotational energy, it is typically assumed to be equilibrated with the translational energy, and therefore is not tracked separately with its own transport equation in CFD codes. To investigate the role of rotational energy, we use the new model to simulate the isothermal ($T$=20,000 K) case shown in Fig. \ref{20k-excite} with specific assumptions. Figure \ref{ROT} shows the simulation results (a) when rotational energy is assumed to be in equilibrium with translational energy, (b) when rotational energy is tracked using the Jeans equation but without removal of internal energy due to dissociation, (c) when rotational energy is tracked but now including rotational energy removal due to dissociation, and (d) when rotational energy is equilibrated with translational energy, however the QSS depletion effects are included in the underlying rotational energy distribution functions. For each of the cases studied, the full model formulation for all vibrational energy mechanisms is used, so as to isolate rotational energy effects only.

As seen in Fig. \ref{ROT}, when translational-rotational equilibrium is assumed including Boltzmann distributions for rotational energy (case (a)), the dissociation rate is significantly higher than the baseline DMS result. When the evolution of rotational energy is tracked using the Jeans equation, but rotational energy is not preferentially removed due to dissociation (case (b)), the result is essentially the same as case (a). Since very little dissociation occurs \emph{during} rotational energy excitation, it may not be necessary to include a separate rotational energy equation, even under this rather extreme nonequilibrium condition. When preferential removal of rotational energy due to dissociation is included in the model (case (c), corresponding to $\langle \epsilon_{rot}^d \rangle$ given by Eq.~\ref{avg_edrot_gen}), the dissociation rate decreases slightly. This is consistent with a lower asymptote for rotational energy compared to case (b). This result indicates that proper removal of rotational energy due to preferential dissociation leads to a small increase in model accuracy. Because the Jeans equation ignores the fact that the average rotational energy relaxation rate is affected by the destruction of high rotational energy molecules, the average rotational energy is about $1500 K \ k_B$ higher relative to the DMS in QSS. Relative to the Jeans equation, incorporating preferential removal of energy results in more accurate prediction of the rotational energy at steady state (less than $500 \ K k_B$), but the dissociation rate is only slightly lower as dissociation depends on the high-energy states which contribute only marginally to average energy. 

Finally, if translational-rotational equilibrium is assumed, including depleted rotational distributions along with no preferential removal of rotational energy due to dissociation (case (d)), Fig. \ref{ROT} shows a larger decrease in the dissociation rate. This result indicates that including only depletion effects (corresponding to QSS dissociation) leads to a noticeable increase in model accuracy. Therefore, compared to use of the full model, a model including only rotational energy depletion effects could be quite accurate. Including preferential rotational energy removal would increase the accuracy slightly, however, tracking rotational energy with a separate transport equation may not be necessary. Of course, the use of a separate rotational energy equation may be useful for hybrid particle-continuum methods \cite{schwartzentruber2006hybrid, schwartzentruber2008multiscale,schwartzentruber2008hybrid,schwartzentruber2007modular}, where model \emph{consistency} is extremely important. 

.

\subsection{Role of Non-Boltzmann Vibrational Energy Distributions}\label{role_nb}

\begin{figure}
\centering 
\subfigure[Overpopulation Effect]
  {
    \includegraphics[width=3.2in]{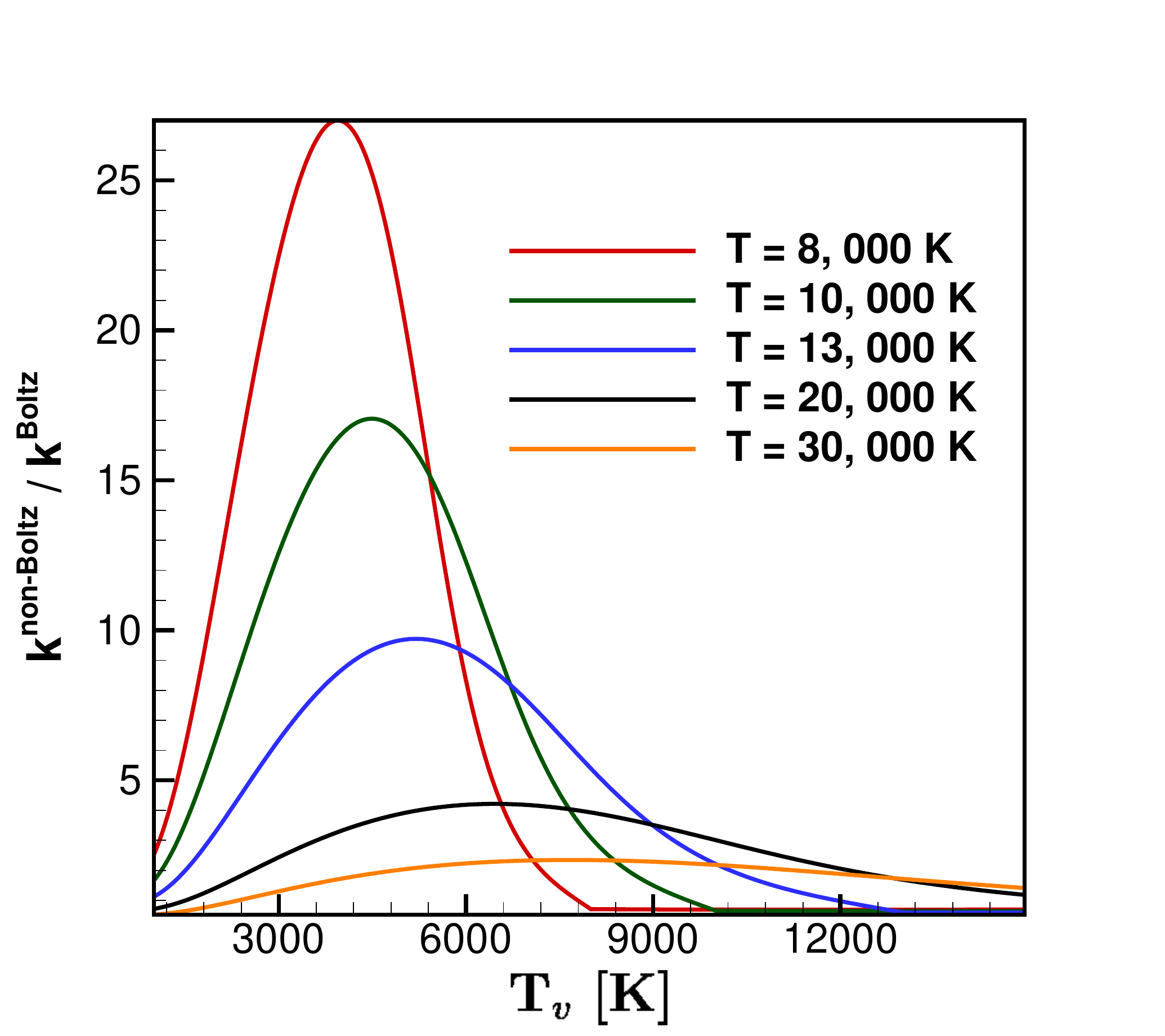}
   \label{overpopulation_transient_ratio}
   }  
   \subfigure[Depletion Effect]
 {
    \includegraphics[width=3.2in]{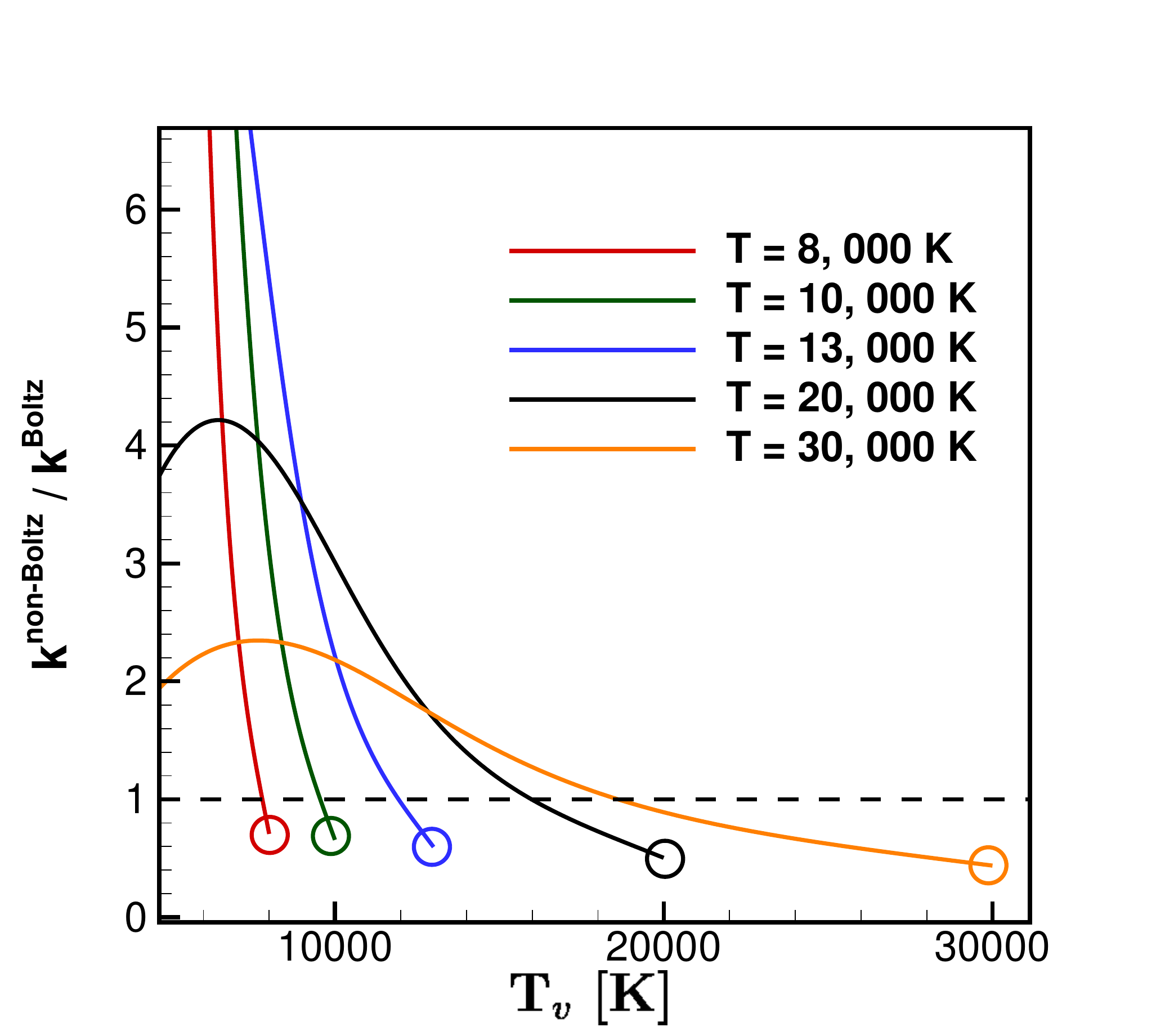}
    \label{depletion_transient_Ratio}
 }  
   \caption{Ratio of non equilibrium reaction rate constant to the rate constant at $T_v$ with Boltzmann distribution in transient phase. The final states denoted by circles (in Fig.~(b)) correspond to $\langle \epsilon_v \rangle  = \langle \epsilon_v \rangle^*$. This figure is also presented in Ref. \cite{singhmodeldevelopment2019}}
   \label{knb_kb_ratio}
\end{figure}

In this subsection, we analyze the relative importance of modeling vibrational nonequilibrium effects. Using the new model, the influence of overpopulation and depletion of high vibrational energy states is quantified in Fig. \ref{knb_kb_ratio}. Specifically, the ratio of the full nonequilibrium rate constant to the equilibrium rate constant is plotted as a function of vibrational temperature at various translational-rotational temperatures. 
In Fig. ~\ref{knb_kb_ratio}, it is evident that overpopulation of high-vibrational levels, in the rapid excitation phase ($T >> T_v$) increases the rate of dissociation (see Fig.~\ref{overpopulation_transient_ratio}). Clearly, overpopulation effects are more prominent at low translational temperature, raising the dissociation rate coefficient by as much as 27x compared to the rate determined using the average vibrational energy (Boltzmann assumption). This trend is consistent with the result that rate constants depend on vibrational temperature, and consequently on the population of high v-states, more strongly at lower translational temperature.

When the gas reaches higher vibrational temperatures, corresponding to the QSS dissociating phase, as shown in Fig.~\ref{depletion_transient_Ratio}, the non-Boltzmann rates are lower due to the depletion of high vibrational energy states. The reduction in the dissociation rate coefficient under such QSS conditions ($T \approx T_v$) is rather constant across a wide range of translational temperatures, a result also found in master-equation and DMS studies \cite{valentini2016dynamics,macdonald2018construction_DMS,grover2019jtht,grover2019direct}.
It should be noted, however, that the extent of depletion and consequent reduction in the dissociation rate will be more prominent in an actual physical ensemble of gas, since the true QSS state has $\langle \epsilon_v \rangle  < \langle \epsilon_v \rangle^*$, which is not the same as the equilibrium condition $\langle \epsilon_v \rangle = \langle \epsilon_v \rangle^*$, plotted in Fig.~\ref{depletion_transient_Ratio}. Finally, although the effects of overpopulation appear dramatic (Fig.~\ref{overpopulation_transient_ratio}), the Boltzmann rate coefficient ($k^{Boltz}$) is low, since the average vibrational energy (represented by $T_v$) is low. Therefore, although overpopulation dramatically increases the dissociation rate relative to the Boltzmann assumption, it is not clear if this will noticeably affect the overall dissociation trend. 

In order to quantify the contribution of vibrational nonequilibrium effects on the overall dissociation process, we use variations of the new model to simulate gas evolution under isothermal conditions and compare to baseline DMS results. 
We first analyze the role of depletion during the QSS phase and, subsequently we analyze over-population effects during the excitation phase. Based on the results of Section 4.1, for the cases simulated in this section, we assume translational-rotational equilibrium and we also neglect preferential removal of rotational energy due to dissociation. However, we do include depletion effects in the rotational energy distribution as this was determined to be the dominant rotational energy mechanism. In terms of the vibrational energy mechanisms, we now present model results for two scenarios, (a) where vibrational energy distributions are assumed to be Boltzmann based on the local average vibrational energy of the gas, and (b) where only depletion effects (not overpopulation) effects are included in the vibrational distributions. Both of these model versions are compared to the full model results, which include both depletion and overpopulation effects.

\begin{figure}
\centering 
\subfigure[]
 {
     \includegraphics[width=3.2in]{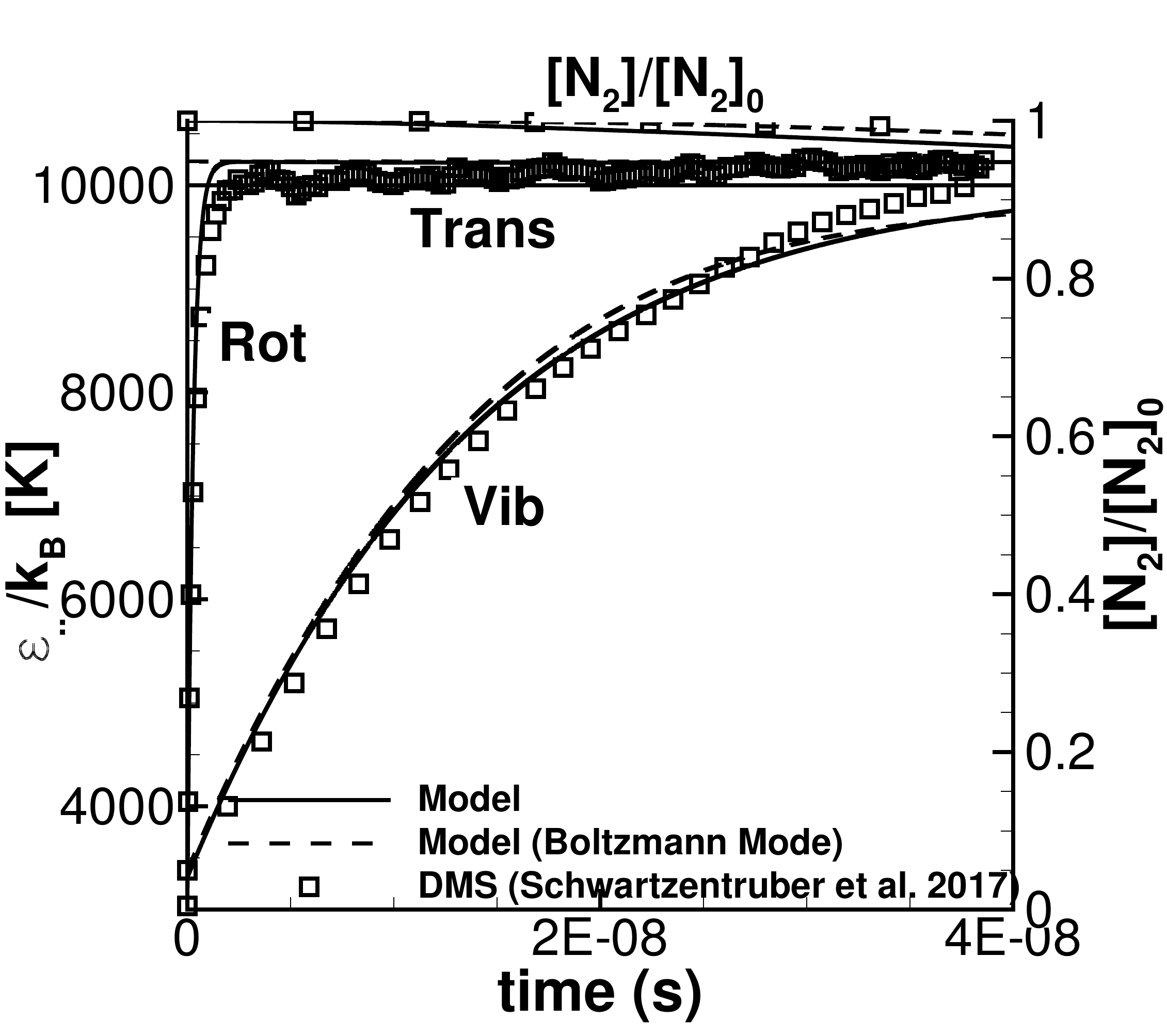}
   \label{10k-excite-VIB}
 } 
 \subfigure[]
  {
    \includegraphics[width=3.2in]{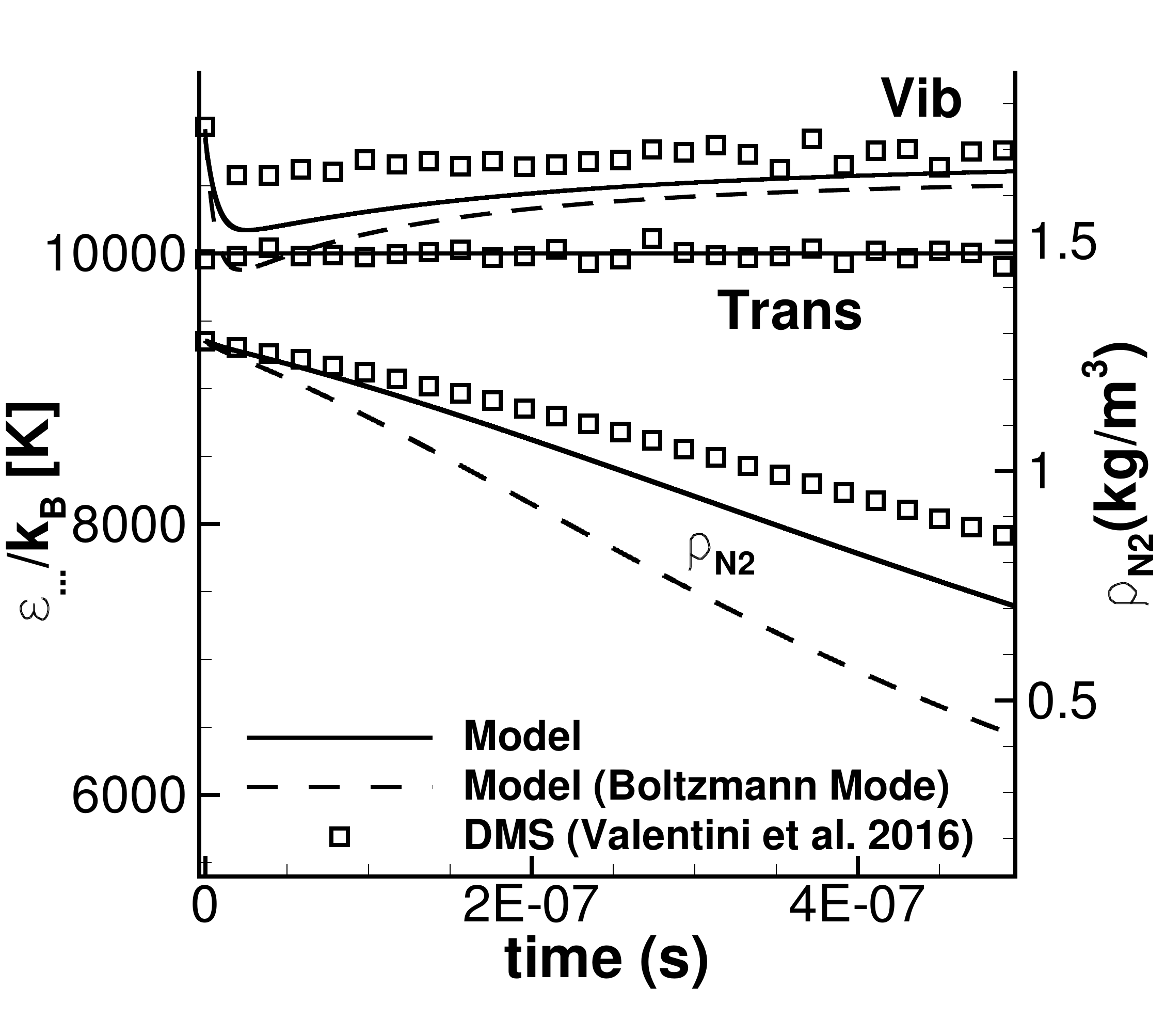}
   \label{10k-dep-VIB}
   }  
   \caption{Isothermal relaxation of nitrogen at $T=10,000$ K analyzing the role of overpopulation and depletion of high energy vibrational states. (a) Excitation case (b) QSS phase }
   \label{overpop-vib}
\end{figure}

\begin{figure}
\centering 
     \includegraphics[width=3.2in]{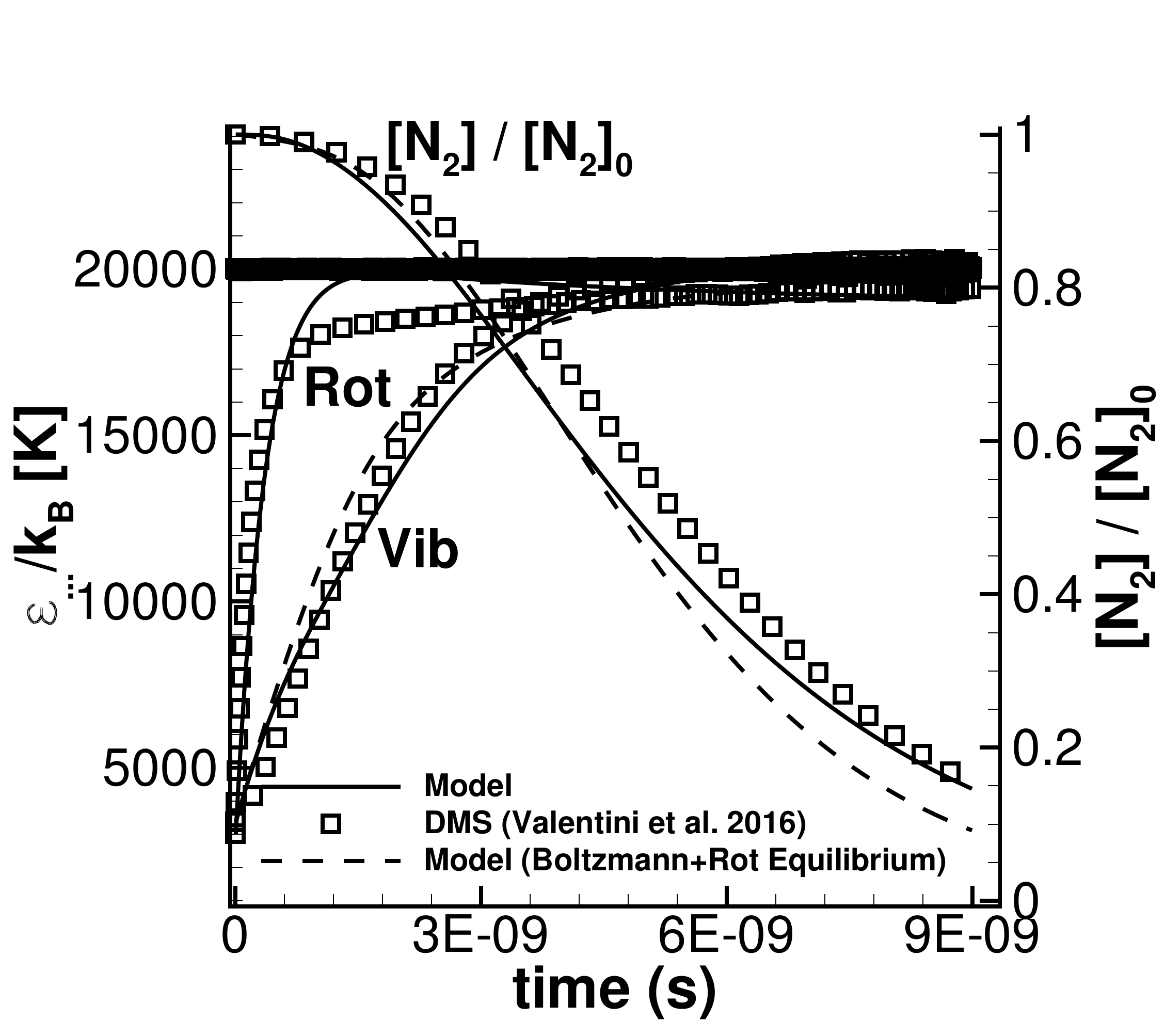}
 \caption{Isothermal ro-vibrational relaxation of nitrogen at $T=20,000$ K to analyze the role of overpopulation and depletion of high vibrational energy states}
  \label{20k-excite-VIB}
\end{figure}

Figure \ref{10k-excite-VIB} shows model results for the isothermal case previously presented in Fig. \ref{10k-excite}. For this case, there is no significant difference between the full nonequilibrium model and the model variant assuming Boltzmann vibrational distributions functions (scenario (a)). This is because very little dissociation occurs during the vibrational excitation phase at a temperature of 10,000 K. Figure \ref{10k-dep-VIB} shows results for the isothermal case previously presented in Fig. \ref{10k-equil}, where all energy modes are initialized to a temperature of 10,000 K. In this case, when vibrational distributions are modeled as Boltzmann (scenario (a)), the resulting dissociation rate is significantly faster than computed by DMS. Since during such QSS dissociation the distributions functions should be depleted (non-Boltzmann). Indeed, the full model includes depletion and is in better agreement with DMS. 

Figure \ref{20k-excite-VIB} shows model results for the isothermal case previously presented in Fig. \ref{20k-excite}. Under the assumption of Boltzmann vibrational energy distributions, the results clearly show an increased rate of dissociation in the QSS phase (i.e. for times greater than $6 \times 10^{-9}$s). By comparing the Boltzmann result with the full model result, which includes both overpopulation and depletion effects, Fig. \ref{20k-excite-VIB} shows that at early times the full model predicts a slightly increased rate of dissociation (due to overpopulation) and then a noticeably slower rate of dissociation during QSS (due to depletion). However, quantitatively, the effect of overpopulation on the overall dissociation process is rather insignificant even under such extreme conditions. This is a result of the fact that although overpopulation can increase the dissociation rate substantially compared to the Boltzmann rate (refer to Fig. \ref{overpopulation_transient_ratio}), the Boltzmann dissociation rate is so low during the vibrational excitation period (due to low average vibrational energy) that overall, there is a negligible effect (for times less than $4 \times 10^{-9}$s in this case). Specifically, there is less than a 5\% difference in N$_2$ concentration between Boltzmann and full model results during the overpopulation phase; a difference that is likely not measurable by experiment. 

In summary, overpopulation effects during the excitation phase are small at high temperatures and do not contribute at low temperatures due to insignificant levels of overall dissociation. In contrast, by comparing Boltzmann and full model results in Fig. \ref{20k-excite-VIB}, including depletion effects clearly increases model accuracy during the QSS phase where most of the dissociation occurs.

\subsection{Role of N$_2$-N$_2$ and N-N$_2$ Collisions}\label{N3_N4}
In this section, we use the new nonequilibrium model to simulate isothermal relaxation and dissociation where only N$_2$-N$_2$ collisions are considered. Specifically, the same conditions as previously presented in Fig. \ref{20k-excite} are simulated, except that when atoms are created due to dissociation, they are immediately removed from the simulation. This enables the model to be compared to previously published DMS calculations that used same approach to isolate processes specific to N$_2$-N$_2$ collisions. Full details of the DMS calculations for this case are presented in Ref. \cite{valentini2015N4}. 

\begin{figure}
\centering 
   \subfigure[Original Model, Park Model and DMS.]
  {
    \includegraphics[width=3.2in]{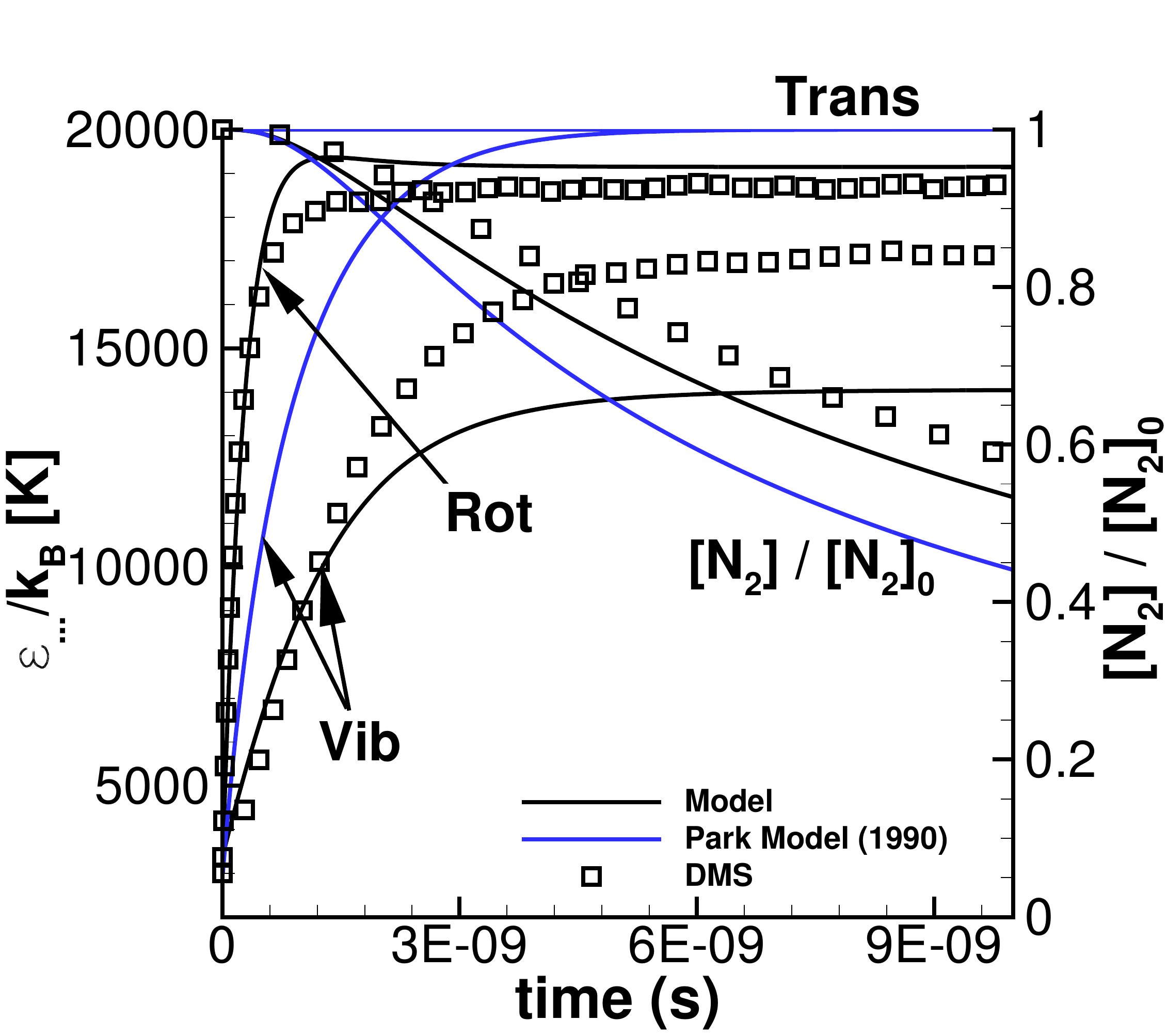}
   \label{zerod_20K_N4}
   }  
   \subfigure[Modified Model and DMS.]
  {
    \includegraphics[width=3.2in]{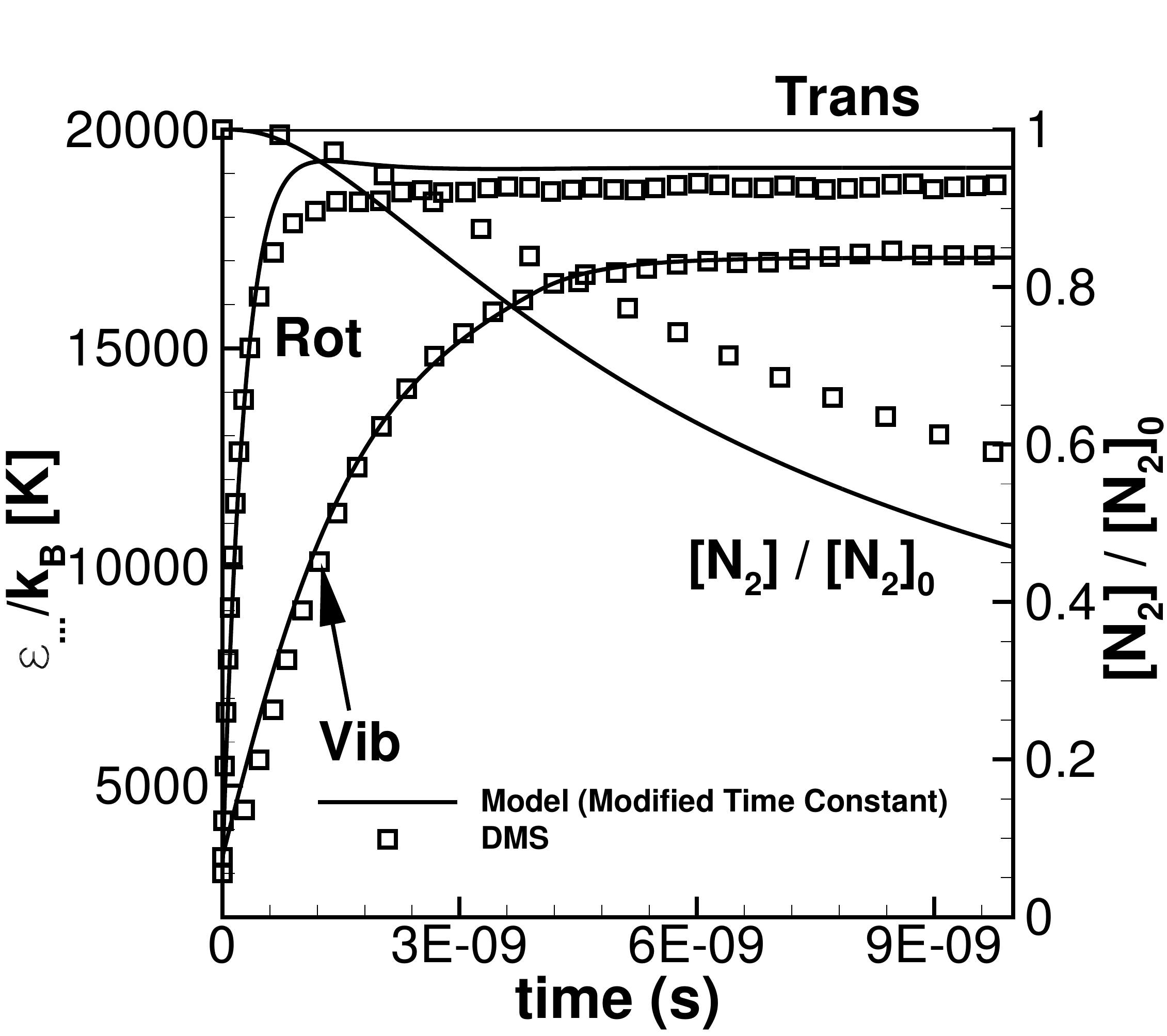}
   \label{zerod_20K_N4_time_constant}
   }  
   
   \caption{Isothermal ro-vibrational relaxation of nitrogen at $T=20, 000$ K, where only N$_2$-N$_2$ interactions are considered. For DMS calculations details please refer to Ref. \cite{valentini2015N4}} 
   \label{zerod_20K_N4-both}
\end{figure}

As shown in Fig.~\ref{zerod_20K_N4}, when only terms corresponding to N$_2$-N$_2$ collisions are kept, the new model agrees fairly well with the DMS data for rovibrational excitation and dissociation, however, doesn't reach precisely the same QSS state (average rotational and vibrational energy) compared to DMS. The Park model shows significant differences compared to DMS with the same level of discrepancy as shown in the previous comparisons in this article. We find the QSS internal energy state to be highly sensitive to model details in such diatom-only calculations. For example, the slightly faster dissociation rate predicted by the new model leads to lower pressure (since atoms are removed), which leads to slower vibrational excitation. As a result, since the QSS state is a balance between internal energy relaxation towards the translational energy and internal energy lost due to dissociation, the precise QSS state is difficult to predict with the new simple model. To show just how sensitive the results are, Fig. \ref{zerod_20K_N4_time_constant} shows the model result when the relaxation time constant ($\tau_v$) is increased by only a factor of 2. Recall from Fig. \ref{time_constant_vib} that $\tau_v$ varies by many orders of magnitude for N$_2$-N$_2$ collisions. This adjustment is enough to raise the QSS average vibrational energy by several thousands of Kelvin, and bring the model prediction into agreement with DMS. 

Although such diatom-only simulations are useful for studying individual contributions to dissociation from both N$_2$-N$_2$ and N-N$_2$ collisions, the resulting QSS  is not physically relevant for most flows. N$_2$-N$_2$ collisions will dominate the excitation and dissociation process immediately behind a shock wave, before the gas is significantly dissociated. However, for flows with substantial dissociation, N-N$_2$ collisions will dominate in the QSS regime. As a result, even though the new model differs from DMS for the condition in Fig. \ref{zerod_20K_N4}, this will have little-to-no effect on model accuracy for a realistic dissociation process. If an application arises where the model accuracy is insufficient, the model could be extended to treat exchange collisions in a manner similar to dissociation reactions, with a separate rate constant. Since exchange collisions are the main reason for the order-of-magnitude difference in relaxation rates between N$_2$-N$_2$ and N-N$_2$ collisions, it is possible that modeling exchange collisions explicitly could increase the fidelity of the model.

\section{Consistent Simplified Non-Equilibrium Models: Rate constant and average vibrational energy of dissociating molecules for CFD}
While the full nonequilibrium model can be readily implemented into multi-temperature CFD codes, as described in previous sections, simplifying assumptions may be possible with negligible loss in solution accuracy. In this section we outline a series of such assumptions, we derive a compact model expression, and compare the expression with previous models from the literature.

\subsection{Derivation of rate constant}{\label{der_simplify}}
Based on the analysis from previous sections, we assume \textbf{(i)} translational-rotational equilibrium (with no preferential removal of rotational energy), but including rotational energy depletion effects, and \textbf{(ii)} vibrational energy distributions where overpopulation is ignored but depletion effects are included. Under these assumptions the dissociation rate constant (refer to the Appendix) is simplified, and the expression was fully derived in Section 4.3 of Ref. \cite{SinghCFDI} as:
\begin{equation}
\begin{split}
   \hat{k}(T, T_{rot},T_{v}) 
  = A T^{\eta} \exp\left[- \frac{\epsilon_d}{k_B T} \right]    *\left[\hat{\Hn}(\epsilon_d,0,1) +  \hat{\Hn}(\epsilon_d^{\mx},\epsilon_d,2) \right]
  \end{split}
  \label{Rate_Final_BB}
\end{equation}
\begin{equation*}
\eta =\alpha-\frac{1}{2}; \hspace{0.35in} A= \frac{1}{S}  \left(\frac{8 k_B }{\pi \mu_C}\right)^{1/2} \pi b_{\max}^2  C_1 \Gamma[1+\alpha] \left(\frac{k_B }{\epsilon_d}\right)^{\alpha-1}
\end{equation*}
\begin{equation}
\begin{split}
    \hat{\Hn}(x,y,n) = \hspace{3.0in}\\ 
    \frac{\exp[(-1)^{n-1}}{ \hat{\Zn}(T_{rot},T_v)} \frac{\exp\left[ x \hat{\zeta}_{rot} \right]  \hat{\gn} (\hat{\zeta}_{vr}) - \exp\left[ y\hat{\zeta}_{rot} \right] \hat{\gn} (\zeta_{v}-y\hat{\zeta}_{rot}/\epsilon_d )}{k_B \theta_{rot} \hat{\zeta}_{rot}}  ;
    \label{Ztvtr}
\end{split}
\end{equation}
where
\begin{equation*}
\begin{split}
\hat{\zeta}_{rot}  = -\frac{1}{k_B T_{rot}}+\frac{1}{k_B T}
+\frac{\beta-\theta_{CB}+(-1)^n\delta}{\epsilon_d} -\frac{\theta_{CB}}{k_B T}+\frac{\hat{\delta}_{rot} }{\theta_{rot} k_B};\hspace{0.10 in}\\
\label{delta_defs}
\end{split}
\end{equation*}
\begin{equation*}
\begin{split}
\zeta_{v}  =  -\frac{1}{k_B T_v} +\frac{1}{k_B T}  +\frac{\gamma+(-1)^n\delta}{\epsilon_d} ; \hspace{0.5in} \hat{\zeta}_{vr} = \zeta_{v}-\hat{\zeta}_{rot}
\label{delta_defs}
\end{split}
\end{equation*}
Here, $\hat{\zeta}_{vr}=\zeta_{v}-\hat{\zeta}_{rot}$. 
Focusing only on the contribution from bound molecules, and assuming that rotational and vibrational energy are independent, the basic dependence of the nonequilibrium term is, 
\begin{equation}
    \hat{\Hn}(\epsilon_d,0,1) \propto  \cfrac{\gn (\zeta_{v})}{-\zeta_{rot}}~.
\end{equation}
Here, $\gn (\zeta_{v})$ can be viewed as a partition function evaluated at an effective vibrational temperature of $T_{\ef}$, where
\begin{equation}
   -\cfrac{1}{k_B T_{\ef}} = -\frac{1}{k_B T_v}+\frac{1}{k_B T} +\frac{\gamma-\delta}{\epsilon_d}~.
   \label{TFp}
\end{equation}
In this case, increasing $T_v$ (and $\gamma$) will increase the rate constant due to a corresponding higher population of $v$-states. Interestingly, the resulting expression resembles the model of Marrone and Treanor \cite{marrone1963chemical}, where the effective temperature in their model is identical to the one given in Eq.~\ref{TFp}. Following the same simplifications, but now for quasi-bound molecules ($\hat{\Hn}(\epsilon_d^{\mx},\epsilon_d,2)$), the effective temperature $T_{\ef}$ according to our model is
\begin{equation}
\begin{split}
       -\cfrac{1}{k_B T_{\ef}} = \hspace{3.0in} \\  -\frac{1}{k_B T_v} + \frac{1}{k_B T_{rot}}
+\frac{\gamma-\beta+\theta_{CB}}{\epsilon_d} + \frac{\theta_{CB}}{k_B T}-\frac{\hat{\delta}_{rot} }{\theta_{rot} k_B}+\frac{\hat{\delta}_{v}}{k_B \theta_{v}^m}~.
\label{TFp-2}
\end{split}
\end{equation}

Recall that $\hat{\Hn}(\epsilon_d,0,1)$ and $\hat{\Hn}(\epsilon_d^{\mx},\epsilon_d,2)$ correspond to contributions from bound and quasi-bound molecules, respectively.  Based on the results of Bender et al. \cite{bender2015improved}, it may be reasonable to assume that \textbf{(iii)} the contribution from bound molecules is a multiple ($C'$) of the contribution from quasi-bound molecules:
\begin{equation}
\hat{\Hn}(\epsilon_d,0,1)  = C' \hat{\Hn}(\epsilon_d^{\mx},\epsilon_d,2)
\end{equation}

The rate expression now contains one nonequilibrium controlling function, $\hat{\Hn}(\epsilon_d^{\mx},\epsilon_d,2)$ and a free parameter ($C'$). Modifying notation for $T\neq T_v$, we write $   \hat{\Hn}(\epsilon_d^{\mx},\epsilon_d,2; T, T_{rot},T_v ) \equiv  \hat{\Hn}(\epsilon_d^{\mx},\epsilon_d,2) , \hat{\gn} (\hat{\zeta}_{vr}; T, T_{rot} T_v ) \equiv \hat{\gn} (\hat{\zeta}_{vr} )$. The expression for $\hat{\Hn}(\epsilon_d^{\mx},\epsilon_d,2)$ is reproduced here from appendix (See Eq.~\ref{HTvtr_rate_QSS})
\begin{widetext}
\begin{equation}
\begin{split}
 \hat{\Hn}(\epsilon_d^{\mx},\epsilon_d,2; T, T_{rot},T_v ) \ = \hspace{3.0in}\\  \frac{\exp[(-1)^{n-1}]}{\hat{\Zn}(T_{rot},T_v)} \frac{\exp\left[ \epsilon_d^{\mx} \hat{\zeta}_{rot}\right] \hat{\gn} (\hat{\zeta}_{vr}) - \exp\left[ \epsilon_d \hat{\zeta}_{rot}\right] \hat{\gn} (\hat{\zeta}_{vr})}{k_B \theta_{rot} \hat{\zeta}_{rot}}  ;
   \\
   =  \frac{\hat{\gn} (\hat{\zeta}_{vr}; T, T_{rot}, T_v)}{\hat{\Zn}(T_{rot},T_v)}  \frac{\exp\left[ \epsilon_d^{\mx} \hat{\zeta}_{rot}\right]   - \exp\left[ \epsilon_d \hat{\zeta}_{rot}\right] }{\exp[1] k_B \theta_{rot} \hat{\zeta}_{rot}}  ;
    \label{Ztvtr}
\end{split}
\end{equation}
\end{widetext}
and, as result of assumption (i), since $T_{rot}\approx T$, the expression can be written as:
 \begin{equation}
\begin{split}
   \hat{\Hn}(\epsilon_d^{\mx},\epsilon_d,2; T, T_{rot},T_v ) = \frac{\hat{\gn} (\hat{\zeta}_{vr}; T, T_{rot}, T_v)}{\hat{\Zn}(T_{rot},T_v)}  \chi(T);
    \label{Ztvtr}
\end{split}
\end{equation}


The next step is to use the function $\hat{\Hn}(...)$, to formulate the \emph{nonequilibrium correction factor}, equal to the ratio of the nonequilibrium rate to the equilibrium rate, denoted by $\hat{\Omega}$: 
\begin{widetext}
\begin{equation}
    \hat{\Omega}(T,T_{rot},T_v)  \equiv \cfrac{\hat{\Hn}(\epsilon_d^{\mx},\epsilon_d,2; T, T_{rot},T_v ) }{\hat{\Hn}(\epsilon_d^{\mx},\epsilon_d,2; T, T,T ) |_{\hat{\delta}_{...}=0}} = \cfrac{\hat{\gn} (\hat{\zeta}_{vr}; T, T_{rot}, T_v)}{\hat{\gn} (\hat{\zeta}_{vr}; T, T, T)|_{\hat{\delta}_{...}=0}} \cfrac{\Zn(T,T)}{\hat{\Zn}(T_{rot},T_v)}
    \\
    = \cfrac{\hat{\gn} (\hat{\zeta}_{vr}; T, T_{rot}, T_v)}{\gn (\zeta_{vr}; T, T, T)} \cfrac{\Zn(T,T)}{\hat{\Zn}(T_{rot},T_v)}
    \\
\end{equation}
\end{widetext}
where removing depletion terms ($\hat{\delta_v} =\hat{\delta}_{rot} =0$), reduces the functions to the Boltzmann distributions based estimates ($\hat{\gn} (\hat{\zeta}_{vr}; T, T_{rot}, T_v)|_{\hat{\delta}_{...}=0} \rightarrow \gn (\zeta_{vr}; T, T_{rot}, T_v)$ and $\hat{\zeta}_{vr}|_{\hat{\delta}_{...}=0} \rightarrow \zeta_{vr} $). The Boltzmann distributions based estimates for the expressions of the quantities such as $\Hn$ and $\gn$ are presented in Sec.~B of Ref.~\cite{SinghCFDI}. 
Since depletion of internal energy states due to dissociation primarily affects the highest energy levels, we further assume that \textbf{(iv)} depletion terms can be omitted from the internal energy partition functions ($\hat{Z}(T_{rot},T_v)\rightarrow Z(T_{rot},T_v)$), reducing the expression to:
\begin{equation}
    \hat{\Omega} (T,T_{rot},T_v)   = \cfrac{\hat{\gn} (\hat{\zeta}_{vr}; T, T_{rot}, T_v)}{\gn (\zeta_{vr}; T, T, T)} \cfrac{\Zn(T,T)}{\Zn(T_{rot},T_v)}
    \label{nb_c_main}
\end{equation}
where the expression for $\hat{\zeta}_{vr}$ is : 
\begin{equation}
    \hat{\zeta}_{vr} = -\frac{1}{k_B T_v} +\frac{1}{k_B T_{rot}}+\frac{B_1}{\epsilon_d} +\frac{\theta_{CB}}{k_B T} -\frac{\hat{\delta}_{rot}}{k_B \theta_{rot}}\ \hspace{0.10 in}\\
\end{equation}
where $B_1 = \gamma-\beta+\theta_{CB} $.

The next set of assumptions deal with simplification of the partition function expressions. Recall that the function $g(x)$ represent the contribution from three different characteristic vibrational temperatures, that was necessary to accurately describe anharmonic characteristics of the diatomic potential energy, and therefore necessary to construct accurate Boltzmann distributions functions (refer to Section 4.1 in Ref. \cite{SinghCFDI}):
\begin{equation*}
  g(x)= \sum_m g_m(x)
\end{equation*}
\begin{widetext}
\begin{equation*}
    g_m(x)=  \exp\left[x E_{m^-}+\hat{\delta}_v m^-\right]\frac{1-\exp\left[ \theta_v^m\ k_B (m^+-m^-)\left(x\ +\cfrac{\hat{\delta}_v}{\theta_v^m\ k_B} \right) \right]}{1-\exp\left[ \theta_v^m\ k_B\left(x\ +\cfrac{\hat{\delta}_v}{\theta_v^m\ k_B} \right) \right]} 
\end{equation*}
\end{widetext}
In order to further simplify, we assume \textbf{(v)} that the partition functions can be approximated by the product of rotational and vibrational distribution functions. This reduces Eq.~\ref{nb_c_main} to:
\begin{equation}
    \hat{\Omega}(T,T_{rot},T_v)   = \cfrac{\hat{\gn} (\hat{\zeta}_{vr}; T, T_{rot}, T_v)}{\gn (\zeta_{vr}; T, T, T)} \cfrac{\Zn(T)}{\Zn (T_v)} \cfrac{T}{T_{rot}^\text{eff}}
    \label{nb_c}
\end{equation}
Finally, we assume that \textbf{(vi)} the partition function contributions can be approximated by the SHO model, and that there exists a single \emph{effective} characteristic temperature $\theta_v^{\text{eff}}$, that can account for anharmonic effects (in place of three separate contributions in $g(x)$). This assumption is reasonable from the standpoint that the low-lying energy states, that contribute significantly to the partition function, are accurately modeled by SHO theory. Furthermore, since a new free parameter ($\theta_v^{\text{eff}}$) is introduced, this parameter may be ``fit'' in order to correct any overall inaccuracy introduced by this assumption. Following this assumption, the SHO partition function and the resulting $g(x)$ function are written as:
\begin{equation}
    Z(T) = \cfrac{1-\exp\left[-\cfrac{\epsilon_d}{k_B T}\right]}{1-\exp\left[-\cfrac{\theta_v}{T}\right]}
\end{equation}
and
\begin{equation}
    g\left(-\cfrac{1}{k_B T}\right) = \cfrac{1-\exp\left[-\cfrac{\epsilon_d}{k_B T} +\cfrac{\hat{\delta}_v \epsilon_d}{\theta_v^{\text{eff}}\ k_B}\right]}{1-\exp\left[-\cfrac{\theta_v^{\text{eff}}}{T} +\hat{\delta}_v\right]}
    \label{g_to_Zeff}
\end{equation}

We can now define a new function (like partition function) which combines T and $\hat{\delta}_v$ terms in Eq.~\ref{g_to_Zeff} as:
\begin{equation}
    Z_\text{eff}(T')= \cfrac{1-\exp\left[-\cfrac{\epsilon_d}{k_B T'} \right]}{1-\exp\left[-\cfrac{\theta_v^{\text{eff}}}{T'} \right]}
\end{equation}
where $T'$ is defined as:
\begin{equation}
    \cfrac{1}{k_B T'} = \cfrac{1}{k_B T} -\cfrac{\hat{\delta}_v }{k_B \theta_v^{ \ \text{eff}}}
\end{equation}

The expression for the nonequilibrium correction factor now reduces to the following compact form:
\begin{equation}
    \hat{\Omega}_{NB}(T,T_{rot},T_v)   = \cfrac{Z_{\text{eff}}(T_{F'})}{Z_{\text{eff}}(-U')} \cfrac{\Zn(T)}{\hat{\Zn}(T_v)} \cfrac{T}{T_{rot}^\text{eff}}
    \label{nb_c}
\end{equation}
where 
\begin{equation}
    T_{rot}^{\text{eff}} = \cfrac{T_{rot}}{1-\cfrac{\hat{\delta}_{rot}}{k_B \theta_{rot}}} 
\end{equation}
\begin{equation}
\begin{split}
   \frac{1}{k_B U'} = \frac{B_1}{\epsilon_d} +\frac{\theta_{CB}}{k_B T}
    \end{split}
\end{equation}
where $\hat{\zeta}_{vr}$ and $\hat{\delta}_v$ terms in Eq.~\ref{nb_c_main}, are combined to define an effective temperature ($T_{F'}$) as:
\begin{equation}
\begin{split}
   \cfrac{1}{k_B T_{F'}} = \frac{1}{k_B T_v} -\frac{1}{k_B T_{rot}}-\frac{B_1}{\epsilon_d} -\frac{\theta_{CB}}{k_B T} +\frac{\hat{\delta}_{rot}}{k_B \theta_{rot}}-\frac{\hat{\delta}_{v}}{k_B \theta_{v}^{\text{eff}}}\ \hspace{0.10 in}\\
    =\frac{1}{k_B T_v} -\frac{1}{k_B T_{rot}}-\frac{1}{k_B U'} +\frac{\hat{\delta}_{rot}}{k_B \theta_{rot}}-\frac{\hat{\delta}_{v}}{k_B \theta_{v}^{\text{eff}}}\ \hspace{0.10 in}
    \end{split}
    \label{TF_current_model}
\end{equation}

Therefore, given the above assumptions \textbf{(i) - (vi)}, the nonequilibrium rate coefficient model is obtained as the product of the equilibrium rate coefficient (standard modified Arrhenius expression based on $T$) and the nonequilibrium correction factor $\hat{\Omega}_{NB}(T,T_{rot},T_v)$ given by Eq. \ref{nb_c}. 
The remaining model parameters are un-changed from the full nonequilibrium model and have physics-based values as listed in the Appendix.

\subsection{Derivation of the average vibrational energy of dissociating molecules}{\label{der_edv_simplify}}
The full general expression for $ \langle \hat{\epsilon}^{d}_v\rangle(T,T_{rot},T_{v})$ is derived by the authors in Ref.~\cite{SinghCFDI}. The expression for $ \langle \hat{\epsilon}^{d}_v\rangle(T,T_{rot},T_{v})$ is:
\begin{equation}
    \langle \hat{\epsilon}^{d}_v\rangle(T,T_{rot},T_{v}) =  \frac{\hat{\Phi}(\epsilon_d,0,1)+\hat{\Phi}(\epsilon_d^{\mx},\epsilon_d,2)}{\hat{\Hn}(\epsilon_d,0,1)+\hat{\Hn}(\epsilon_d^{\mx},\epsilon_d,2)}
    \label{avg_edv_QSS_simple_1}
\end{equation}
The algebraic steps in the simplification of $ \langle \hat{\epsilon}^{d}_v\rangle(T,T_{rot},T_{v})$ in  Eq.~\ref{avg_edv_QSS_simple_1} are similar to those used in the rate constant.
Therefore, using the same set of assumptions (\textbf{(i) - (vi)}), one can find a simple expression for the average vibrational energy of the dissociating molecules as 
\begin{equation}
\begin{split}
    \langle \hat{\epsilon}^{d}_v\rangle(T,T_{rot},T_{v})    =   E_{0} - \frac{\epsilon_d }{\exp\left[\cfrac{\epsilon_d}{k_B T_F'} \right]-1} 
   + \frac{ \theta_v^{\ef}\ k_B }{\exp\left[ \cfrac{\theta_v^{\ef}}{T_F'}  \right]-1}
    \end{split}
    \label{avg_edv_QSS_simplified}
\end{equation}
where $E_0$ is zero-point energy of the molecule. The derivation details for Eq.~\ref{avg_edv_QSS_simplified} are presented in Sec.\ref{simplify_edv_append} of the appendix.

\subsection{Relation to Prior Dissociation Models}

To model the coupling between vibrational energy and dissociation, Marrone and Treanor \cite{marrone1963chemical} proposed the probability of dissociation as $p = \exp\left[-(\epsilon_d -\epsilon_v)/k_B U\right] $, where $U$ is a free parameter. 
Although Marrone and Treanor \cite{marrone1963chemical} noted that the high vibrational levels are strongly favored for dissociation, the exponential form of the dependence on vibrational energy chosen had no bearing on either theoretical or experimental data. In-fact, Marrone and Treanor \cite{marrone1963chemical} in developing their model mentioned (on page 1216) \textit{``There is a good reason to believe, both theoretically and experimentally, that the dissociation probability is higher for higher vibrational levels. For the calculations presented in this report an exponential distribution of probabilities is used. The exact form of the proper probability distribution has not been determined, although several approaches to this problem have been suggested. The exponential distribution used in the present work allows for rapid solution of the equations, and also weights the upper levels sufficiently to provide some insight into this effect on the computed dissociation rate. Thus, the distribution makes it possible ...."}.

This probability expression analytically leads to an expression for the nonequilibrium correction factor:
\begin{equation}
    \begin{split}
        \Omega_{CVDV} =\cfrac{Z(T)Z(T_F)}{Z(T_v)Z(-U)} 
    \end{split}
\end{equation}
where 
\begin{equation}
     \frac{1}{T_F} = \frac{1}{T_v}-\frac{1}{T}-\frac{1}{U}
\end{equation}

In our dissociation model, if depletion effects are ignored ($\delta_{...}^{NB} =0$), centrifugal barrier effects are ignored ($\theta_{CB}=0$), and the joint ro-vibrational distribution function is expressed as a product of separate rotational and vibrational partition functions, the nonequilibrium correction factor for our model (Eq. \ref{nb_c}) becomes,
\begin{equation}
    \Omega_{NB}(T,T_{rot},T_v) = \cfrac{Z(T)Z(T_F)}{Z(T_v)Z(-U)}~,
\end{equation}
which is identical to the correction from Marrone and Treanor \cite{marrone1963chemical}, where $U$ is linked to $B_1$ as $1/U= B_1 k_B/\epsilon_d$~. The most important reason for this equivalence is that fact that Morrone and Treanor modeled the probability of dissociation as an \emph{exponential} function of the molecule's vibrational energy. Such a dependence is now supported by an enormous amount of ab-intio data from QCT and DMS calculations, which provided the basis for our proposed dissociation model (refer to Fig. 2 and Term-3 in Eq. 10 in Ref.~\cite{SinghCFDI}).

In-fact, the above model for Morrone and Treanor was extended by Knab et al. \cite{knab1995theory} assuming an effective dissociation energy, which was further modified to include rotational energy in Ref.~\cite{kanne1996influence}.
Other researchers have proposed various functional forms for the parameter $U$, including a parametrized model from Andrienko and Boyd \cite{andrienko2015high} and a more generalized expansion from Kustova \textit{et. al.} \cite{kustova2016advanced}:
\begin{equation}
    U(\langle \epsilon_v \rangle,T) = \sum_{n=0}^{N} a_n \langle \epsilon_v \rangle ^n \exp\left[T \sum_{n=0}^{K} b_k \langle \epsilon_v\rangle^k \right]
\end{equation}
Such model formulations could easily be extended to incorporate non-Boltzmann effects, by including similar terms as derived in our model (such as $\hat{\delta_v}$ and $\hat{\delta}_{rot}$ in Eq.~\ref{TF_current_model}). The foregoing discussion is also valid in the context of the simplified expression (Eq.~\ref{avg_edv_QSS_simplified}) for the average vibrational energy of dissociating molecules.

Recently, Chaudhry and Candler \cite{chaudhry2019statistical} have proposed modifications to the parameter $U$ based on ab-intio QCT and DMS results, that are consistent with the analytically derived model presented in this article. This Modified Morrone and Treanor (MMT) model (along with a version of the full nonequilibrium model listed in the Appendix) has been implemented in the US3D CFD code \cite{ChaudhryBTSC2020}. It is important to note that such model simplification often introduces a small number of free parameters, that can be ``tuned'' to obtain very close agreement with overall trends from DMS; potentially obtaining closer agreement than the analytically derived model presented in this article. This is a very pragmatic approach to obtaining accurate, yet efficient, nonequilibrium models for use in large-scale (i.e. billion-element) CFD simulations. Furthermore, models with simplified functional forms are more easily combined with additional (and necessary) physics models, such as electronic energy effects and recombination chemistry \cite{ChaudhryBTSC2020}.

\section{Recombination }
In this article, recombination is not incorporated, however, one can easily modify the evolution equation for vibrational energy (and rotational energy) in the following manner:
\begin{equation}
\begin{split}
    \frac{ d \langle \epsilon_v \rangle}{dt} = \frac{\langle \epsilon_v^* \rangle-\langle  \epsilon_v \rangle}{\tau_{\text{mix}}} - k_{N_2-X} [X] (\langle \epsilon_v^d \rangle - \langle \epsilon_v \rangle) \\ + k_{N_2-X}^r [X] (\langle \epsilon_v^{rec} \rangle  - \langle \epsilon_v \rangle)
    \end{split}
     \label{LandauTeller_modified_general_recomb}
 \end{equation}
 
 \begin{equation}
 \begin{split}
    \frac{ d \langle \epsilon_{rot} \rangle}{dt} = \frac{\langle \epsilon_{rot}^* \rangle-\langle  \epsilon_{rot} \rangle}{\tau_{\text{mix}}} - k_{N_2-X} [X] (\langle \epsilon_{rot}^d \rangle - \langle \epsilon_{rot} \rangle) \\ + k_{N_2-X}^r [X] (\langle \epsilon_{rot}^d \rangle  - \langle \epsilon_{rot} \rangle)
    \end{split}
     \label{Jeans_modified_general_recomb}
 \end{equation}
 where $X$ is the collision partner, which can be $N_2$ or $N$. The species source term is then modified as:
  \begin{equation}
    \frac{ d [N_2]}{dt} = -k_{N_2-X} [N_2][X]+k_{N_2-X}^r [N][N][X]
     \label{Rate_eqn_general_recomb}
 \end{equation}
 where using detailed balance
 \begin{equation}
     K_c = \cfrac{[N]^*[N]^*}{[N_2]^*}. 
     \label{equilibrium_constant}
 \end{equation}
where $k_{N_2-X}^r$ is the recmbination rate constant and $\langle \epsilon_v^{rec} \rangle$ is the average vibrational energy of the molecules formed via recombination of atoms. For the recombination rate and average vibrational energy of the molecules formed, simple approximations such as $k_{N_2-X}^r = k_{N_2-X}/K_c$,  and $\langle \epsilon_v^{rec} \rangle =\langle \epsilon_v^d \rangle $ can be imposed.  
Here, the equilibrium constant, $K_c$, can be expressed in terms of the partition function at equilibrium. Note that we have used detailed balance in deriving the recombination rate. However, using microscopic reversibility, the continuum expressions for $k_{N_2-X}^r$ and $\langle \epsilon_v^{rec} \rangle$ from kinetic dissociation rates (derived in Ref.~\cite{SinghCFDI}) can also be derived. To keep the focus of the current article on the dissociation, we intend to incorporate recombination reaction rates including the derivation and implementation details in future studies.


Furthermore, the effect on non-Boltzmann distributions in expanding flows, dominated by $v-v$ transitions and recombination reactions, have not been taken into account in Eqs.~\ref{LandauTeller_modified_general_recomb}-\ref{Rate_eqn_general_recomb}. While recent QCT and DMS studies have provided ab-inito results quantifying the ro-vibrational states that lead to dissociation, there is no such data that quantifies the ro-vibrational states that are populated during recombination. Such a study is beyond the scope of the current research and therefore, at this time, we recommend combining our new dissociation model with the approach for recombination given by Eqs.~\ref{LandauTeller_modified_general_recomb}-\ref{equilibrium_constant}.

\section{Conclusions}

A new nonequillibrium dissociation model and accompanying internal energy excitation models is analyzed. The model was formulated using ab-intio results based on an accurate PES for nitrogen collisions. In Part I (Ref.~\cite{SinghCFDI}), the model was formulated at the kinetic (cross-section) level and enabled analytical integration to obtain a closed-form continuum-level dissociation model. In Part II (this article), the new model is shown to accurately reproduce baseline DMS calculation results for isothermal and adiabatic conditions relevant to strong shock waves induced by hypersonic flight. The widely used Park $TT_v$ model exhibits substantial discrepancies compared to the DMS results (effective dissociation rates 1.5-3.5 times higher than DMS results). 

Since the new model is analytically derived, contributions from each physical mechanism, towards the overall dissociation process, are contained in separate terms that can be individually analyzed. The model includes effects due to average translational, rotational, and vibrational energy, overpopulation and depletion of high internal energy states (non-Boltzmann effects), anharmonic effects in the diatomic potential energy surface, contributions from both bound and quasi-bound molecules, and centrifugal barrier effects. By analyzing the contributions of each mechanism, we make a number of general conclusions. First, it is not necessary to include a rotational energy transport equation, rather, assuming translational-rotational equilibrium is an accurate assumption providing depletion of high rotational energy states (due to dissociation) is included. Second, it is not necessary to model the overpopulation of high vibrational energy states since little dissociation occurs during this phase where vibrational energy is still rapidly exciting. However, depletion of high vibrational energy states due to dissociation must be included for model accuracy. We find the dominant mechanism is that dissociation is exponentially related to vibrational energy. This means that the Boltzmann distribution function must be accurately constructed including anharmonic effects of the PES. Crucially, this also requires preferential removal of vibrational energy due to dissociation reactions to be modeled accurately. Finally, we find that the modified Landau-Teller model is able to accurately model the excitation of vibrational energy, circumventing the need to prescribe a large number of state-specific transition rates using master equation calculations \cite{macdonald2018construction_DMS}.

Based on the above conclusions regarding the relative importance of each mechanism, a series of assumptions are proposed that significantly simplify the functional form of the new nonequilibrium model. The resulting functional form is similar to that proposed by Morrone and Treanor. Both the full nonequilibrium model and various simplified versions can be easily incorporated into state-of-the-art, multi-temperature CFD codes, as a replacement for current empirical models \cite{ChaudhryBTSC2020}.

While simplified model formulations can be made highly accurate by tuning a small number of free parameters to best match ab-initio data, such expressions will no longer be analytically consistent with the underlying kinetic (cross-section based) models \cite{SinghCFDI}. Since the full nonequilibrium model proposed and analyzed in this article is analytically derived it will be useful for hybrid particle-continuum methods, where model \emph{consistency} is crucial.

The main purpose of the work presented in this article is to derive, from first-principles, a dissociation model that accurately captures all the relevant physics and is consistent across kinetic to continuum regimes. Such a model formulation enables the analysis and understanding of each physical mechanism that contributes to nonequilibrium dissociation and enables physics-based model reduction leading to a new multi-temperautre framework that can hopefully replace empirical models.



\section*{Acknowledgement}
This work was supported by Air Force Office of Sci-
entific Research Grants FA9550-16-1-0161 and FA9550-19-1-0219 and was also partially supported
by Air Force Research Laboratory Grant FA9453-17-2-0081. Narendra Singh was partially supported by Doctoral Dissertation Fellowship. Authors are thankful to Dr. E. Torres for providing new DMS results and insights. Furthermore, discussions with Dr. R. Chaudhry, Dr. P. Valentini and Prof. G Candler are greatly acknowledged.

\appendix
\vspace{2in}
\section*{Appendix}
The three inputs required for the zero-dimensional continuum relaxation calculations (Eqs.\ref{Rate_eqn_pop}--\ref{total_energy}) are presented in this appendix; namely, the rate constant ($ k^d_{N_2-X}$ ),  the vibrational-relaxation time constant ($\tau_{v,X}$) and the average energy of the dissociating molecules ($\langle \epsilon_v^d \rangle$) are presented in this appendix.

\section{Park model \cite{park1989assessment} and Millikan and White correlation \cite{millikan1963systematics}} \label{Existing_CFD}

\begin{enumerate}
  \item  For the dissociation rate constant, Park's two temperature model \cite{park1989assessment,park1993review} is used as follows:
\begin{equation}
    k^d_{N_2-X} = C_x T_{\text{eff}}^\eta \exp\left[-\cfrac{\theta_d}{T_{\text{eff}}} \right]
\end{equation}
where $C_{N_2} = 0.01162 $ cm$^3$molecule$^{-1}$s$^{-1}$, $C_{N} = 0.0498$ cm$^3$molecule$^{-1}$s$^{-1}$, $\eta = -1.6$, $\theta_d =113, 200$ K and $T_{\text{eff}} = \sqrt{T_v T}$.
\item Time constants used in the existing CFD formulation  are based on Millikan and White experimental fits \cite{millikan1963systematics} along with the Park high temperature correction \cite{park1993review}: 
\begin{equation}
    p \tau_{v,X} = \exp \left[a_m \left(T^{-1/3}-b_m\right)-18.42 \right]
\end{equation}
where $ p \tau_{v,X}$ is in atm-s, $p$ is pressure, $a_{N_2} = 221, a_{N} = 180, b_{N_2} = 0.0290, b_{N} =0.0262$.
\item  The average energy of dissociating molecules, $\langle \epsilon_v^d \rangle$ is set as $\langle \epsilon_v \rangle$.
\end{enumerate}

\section{New Ab Initio Model} \label{Abinitio_Model}

\subsection{Dissociation Rate Constant}
The full model expression for the new non-equilibrium model was derived in Sec.~D of Ref.~\cite{SinghCFDII}, and is listed again here for completeness:
\begin{equation}
     k^{NB} = \cfrac{  \tilde{k}(T,T_{rot},T_v; T_0) +\Lambda \hat{k}(T,T_{rot},T) }  {1 + \Lambda } ~.
     \label{rate_full_nb}
\end{equation}
Here, $\tilde{k}(T,T_{rot},T_v; T_0) $ represents the contribution predominantly from the low energy states, and $\hat{k}(T,T_{rot},T)$ represents the contribution from the high energy tail of the distribution. $\Lambda$ controls the relative importance of each term capturing the overpopulation and depletion effects. 
\begin{widetext}
\subsubsection{Contribution from the low-energy states $ \tilde{k}(T,T_{rot},T_v; T_0)$}
The expression for $\tilde{k}(T,T_{rot},T_v; T_0)$ is:
\begin{equation}
\begin{split}
 \tilde{k}(T,T_{rot},T_v; T_0) 
  = A T^{\eta} \exp\left[- \frac{\epsilon_d}{k_B T} \right]    *\left[\tilde{\Hn}(\epsilon_d,0,1) +  \tilde{\Hn}(\epsilon_d^{\mx},\epsilon_d,2) \right]
  \end{split}
  \label{Rate_Final__derive}
\end{equation}

\begin{equation*}
\eta =\alpha-\frac{1}{2}; \hspace{0.35in} A= \frac{1}{S}  \left(\frac{8 k_B }{\pi \mu_C}\right)^{1/2} \pi b_{\max}^2  C_1 \Gamma[1+\alpha] \left(\frac{k_B }{\epsilon_d}\right)^{\alpha-1}
\end{equation*}

\begin{equation}
\begin{split}
    \tilde{\Hn}(x,y,n) = \hspace{3.0in} \\ \frac{\exp[(-1)^{n-1}\delta]}{ \tilde{\Zn}(T_v,T_0,T_{rot})} \frac{\exp\left[ x \hat{\zeta}_{rot} \right]  \tilde{\gn} (\hat{\zeta}_{vr}) - \exp\left[ y\hat{\zeta}_{rot} \right] \tilde{\gn} (\zeta_{v}-y\hat{\zeta}_{rot}/\epsilon_d )}{k_B \theta_{rot} \hat{\zeta}_{rot}}  ;
    \label{Ztvtr_rate__derive}
\end{split}
\end{equation}

where
\begin{equation*}
\begin{split}
\hat{\zeta}_{rot}  = -\frac{1}{k_B T_{rot}}+\frac{1}{k_B T}
+\frac{\beta-\theta_{CB}+(-1)^n\delta}{\epsilon_d} -\frac{\theta_{CB}}{k_B T}+\frac{\hat{\delta}_{rot} }{\theta_{rot} k_B};\hspace{0.10 in}\\
\label{deltarot_defs_QSS}
\end{split}
\end{equation*}
\begin{equation*}
\begin{split}
\zeta_{v}  =  +\frac{1}{k_B T_0} +\frac{1}{k_B T}  +\frac{\gamma+(-1)^n\delta}{\epsilon_d} ; \hspace{0.5in} \hat{\zeta}_{vr} = \zeta_{v}-\hat{\zeta}_{rot}
\label{deltav_defs_QSS}
\end{split}
\end{equation*}

\begin{equation*}
     \begin{split}
         \Delta_{\epsilon} = \epsilon_v(1)-\epsilon_v(0) = k_B \theta_v^I
     \end{split}
 \end{equation*}

The expression for $\tilde{g}$ is:

 \begin{equation}
\begin{split}
    \tilde{\Zn}(T_v,T_0,T_{rot}) = \cfrac{T_{rot}}{\theta_{rot} -\cfrac{ T_{rot}}{\theta_{rot}k_B}\hat{\delta}_{rot} }
    \left \{ \tilde{\gn} \left(+\frac{1}{ k_B T_0}\right)  - \exp\left[-\cfrac{\epsilon_d^{\mx}}{k_B T_{rot}} +\epsilon_d^{\mx} \cfrac{\hat{\delta}_{rot}}{k_B \theta_{rot}} \right] \right. \\ \left. \times \tilde{\gn} \left(+\frac{1}{ k_B T_0}+ \cfrac{1}{k_B T_{rot}} -\cfrac{\hat{\delta}_{rot}}{k_B \theta_{rot}}\right)\right \} ;
    \label{Ztvtr_qss_derive}
\end{split}
\end{equation}

\begin{equation*}
    \hat{\delta}_v =  - \lambda_{1,v} \frac{  3 k_B T }{2\epsilon_d}  \hspace{0.25in}  \hat{\delta}_{rot}=  - \lambda_{1,j} \frac{  3 k_B T }{2\epsilon_d} \hspace{0.25in} 
\end{equation*}

where,
\begin{equation*}
  \tilde{\gn}(x)= \sum_m \tilde{\gn}_m(x)
\end{equation*}

\begin{equation*}
    \tilde{\gn}_m(x)=  \exp\left[x E_{m^-}+ \tilde{\delta}_v m^-\right]\frac{1-\exp[ (m^+-m^-) (x \theta_v^m\ k_B+ \tilde{\delta}_v) ]}{1-\exp[x\  \theta_v^m\ k_B+ \tilde{\delta}_v ]} 
\end{equation*}

\begin{equation*}
   \tilde{\delta}_v =  \hat{\delta}_v  -\cfrac{\Delta_{\epsilon} }{k_B T_v}-\cfrac{\Delta_{\epsilon} }{k_B T_0} \hspace{0.25in} 
\end{equation*}

and where $\hat{\delta}_v$ and $\hat{\delta}_{rot}$ accounts for the depletion in the population due to dissociation.

The expression for the derivative of $\tilde{\gn}(x)$ denoted as $\tilde{g}'(x)$ is:
\begin{equation*}
   \tilde{\gn}'(x)= \sum_m \tilde{\gn}'_m(x)
\end{equation*}
\begin{equation*}
    \tilde{\gn}'_m(x)= \frac{\partial \tilde{\gn}_m}{\partial x} = \tilde{\gn}_m\frac{\partial \log \tilde{\gn}_m}{\partial x} 
\end{equation*}

\begin{equation*}
\begin{split}
   \tilde{\gn}'_m(x)=\tilde{\gn}_m(x)\left\{ E_{m^-} - \frac{(m^+-m^-)  \theta_v^m\ k_B \exp[ (m^+-m^-) (x \theta_v^m\ k_B +\tilde{\delta}_v )]}{1-\exp[ (m^+-m^-)( x\theta_v^m\ k_B +\tilde{\delta}_v)]} \right. \\
   \left.
   + \frac{ \theta_v^m\ k_B \exp[x\  \theta_v^m\ k_B +\tilde{\delta}_v ]}{1-\exp[x\  \theta_v^m\ k_B +\tilde{\delta}_v  ]} \right \} 
\end{split}
\end{equation*}

\subsubsection{ Contribution from the high-energy states, $\hat{k}(T, T_{rot},T_{v})$ }
The expression for $\hat{k}(T, T_{rot},T_{v})$ is:
\begin{equation}
\begin{split}
   \hat{k}(T, T_{rot},T_{v}) 
  = A T^{\eta} \exp\left[- \frac{\epsilon_d}{k_B T} \right]    *\left[\hat{\Hn}(\epsilon_d,0,1) +  \hat{\Hn}(\epsilon_d^{\mx},\epsilon_d,2) \right]
  \end{split}
  \label{Rate_Final_QSS}
\end{equation}
\begin{equation*}
\eta =\alpha-\frac{1}{2}; \hspace{0.35in} A= \frac{1}{S}  \left(\frac{8 k_B }{\pi \mu_C}\right)^{1/2} \pi b_{\max}^2  C_1 \Gamma[1+\alpha] \left(\frac{k_B }{\epsilon_d}\right)^{\alpha-1}
\end{equation*}
\begin{equation}
\begin{split}
    \hat{\Hn}(\epsilon_i,\epsilon_j,n) = \hspace{3.0in} \\ \frac{\exp[(-1)^{n-1}\delta]}{ \hat{\Zn}(T_{rot},T_v)} \frac{\exp\left[ \epsilon_i \hat{\zeta}_{rot} \right]  \hat{\gn} (\hat{\zeta}_{vr}) - \exp\left[ \epsilon_j \hat{\zeta}_{rot} \right] \hat{\gn} (\zeta_{v}-\epsilon_j \hat{\zeta}_{rot}/\epsilon_d )}{k_B \theta_{rot} \hat{\zeta}_{rot}}~,
    \label{HTvtr_rate_QSS}
\end{split}
\end{equation}
where,
\begin{equation*}
\begin{split}
\hat{\zeta}_{rot}  = -\frac{1}{k_B T_{rot}}+\frac{1}{k_B T}
+\frac{\beta-\theta_{CB}+(-1)^n\delta}{\epsilon_d} -\frac{\theta_{CB}}{k_B T}+\frac{\hat{\delta}_{rot} }{\theta_{rot} k_B};\hspace{0.10 in}\\
\label{deltarot_defs_QSS}
\end{split}
\end{equation*}
\begin{equation*}
\begin{split}
\zeta_{v}  =  -\frac{1}{k_B T_v} +\frac{1}{k_B T}  +\frac{\gamma+(-1)^n\delta}{\epsilon_d} ; \hspace{0.5in} \hat{\zeta}_{vr} = \zeta_{v}-\hat{\zeta}_{rot}
\label{deltav_defs_QSS}
\end{split}
\end{equation*}

The expression for $\hat{g}$ is:

\begin{equation}
\begin{split}
    \hat{\Zn}(T_{rot},T_{v}) = \cfrac{T_{rot}}{\theta_{rot} -\cfrac{ T_{rot}}{\theta_{rot}k_B}\hat{\delta}_{rot} }
    \left \{ \hat{\gn} \left(-\frac{1}{ k_B T_v}\right)  - \exp\left[-\cfrac{\epsilon_d^{\mx}}{k_B T_{rot}} +\epsilon_d^{\mx} \cfrac{\hat{\delta}_{rot}}{k_B \theta_{rot}} \right] \right. \\ \left. \times \hat{\gn} \left(-\frac{1}{ k_B T_v}+ \cfrac{1}{k_B T_{rot}} -\cfrac{\hat{\delta}_{rot}}{k_B \theta_{rot}}\right)\right \} ;
    \\
    \label{Ztvtr_qss}
\end{split}
\end{equation}
\begin{equation*}
    \hat{\delta}_v =  - \lambda_{1,v} \frac{  3 k_B T }{2\epsilon_d}  \hspace{0.25in}  \hat{\delta}_{rot}=  - \lambda_{1,j} \frac{  3 k_B T }{2\epsilon_d} \hspace{0.25in} ~,
\end{equation*}
where,
\begin{equation*}
  \hat{\gn}(x)= \sum_m \hat{\gn}_m(x)
\end{equation*}
\begin{equation*}
    \hat{\gn}_m(x)=  \exp\left[x E_{m^-}+ \hat{\delta}_v m^-\right]\frac{1-\exp[ (m^+-m^-) (x \theta_v^m\ k_B+ \hat{\delta}_v) ]}{1-\exp[x\  \theta_v^m\ k_B+ \hat{\delta}_v ]} 
\end{equation*}

\subsection{ Average vibrational ($\langle \epsilon^{d}_v \rangle$) and rotational ($\langle \epsilon^{d}_{rot} \rangle$) energy of dissociating molecules} {\label{avgedvrotappend}}
The average vibrational energy of dissociated molecules is calculated in the manner analogous to the rate constant :
\begin{equation}
   \langle \epsilon^{d}_v \rangle ^{NB}(T,T_{rot},T_{v}) =  \cfrac{\langle \tilde{\epsilon}^{d}_v \rangle(T,T_{rot},T_v; T_0) + \langle \hat{\epsilon}^{d}_v\rangle(T,T_{rot},T_{v}) \Lambda  k_r}{1+\Lambda  k_r}
    \label{edv_gen_sum_NB}
\end{equation}
where 
\begin{equation}
    k_r =\cfrac{\hat{k}(T,T_{rot},T) }{\tilde{k}(T,T_{rot},T_v; T_0)}
\end{equation}
and where the expression for $\tilde{\epsilon}^{d}_v(T,T_{rot},T_v; T_0)$ is:

\begin{equation}
    \langle \tilde{\epsilon}^{d}_v\rangle(T,T_{rot},T_{v}) =  \frac{\tilde{\Phi}(\epsilon_d,0,1)+\tilde{\Phi}(\epsilon_d^{\mx},\epsilon_d,2)}{\tilde{\Hn}(\epsilon_d,0,1)+\tilde{\Hn}(\epsilon_d^{\mx},\epsilon_d,2)}
    \label{edv_gen_sum_QSS}
\end{equation}
The functions in Eq.~\ref{edv_gen_sum_QSS} are given by:
\begin{equation*}
\begin{split}
    \tilde{\Phi}(\epsilon_i,\epsilon_j,n) = \cfrac{\exp[(-1)^{n-1}\delta]}{ \tilde{\Zn}(T_v,T_0,T_{rot})} \frac{\exp\left[ \epsilon_i \hat{\zeta}_{rot}\right]  \tilde{\gn}' (\hat{\zeta}_{vr}) - \exp\left[ \epsilon_j \hat{\zeta}_{rot}\right] \tilde{\gn}' (\zeta_{v}-\epsilon_j \hat{\zeta}_{rot}/\epsilon_d )}{k_B \theta_{rot}\hat{\zeta}_{rot}} ;
    \label{Hnb_QSS}
\end{split}
\end{equation*}
\begin{equation*}
\begin{split}
    \tilde{\Hn}(\epsilon_i,\epsilon_j,n) =  \cfrac{\exp[(-1)^{n-1}\delta]}{ \tilde{\Zn}(T_v,T_0,T_{rot})}  \frac{\exp\left[ \epsilon_i \hat{\zeta}_{rot}\right]  \tilde{\gn} (\hat{\zeta}_{vr}) - \exp\left[ \epsilon_j \hat{\zeta}_{rot}\right] \tilde{\gn} (\zeta_{v}-\epsilon_j \zeta_{rot}/\epsilon_d )}{k_B \theta_{rot}\hat{\zeta}_{rot}} ;
    \label{Ztvtr_QSS}
\end{split}
\end{equation*}

and the expression for $\langle \hat{\epsilon}^{d}_v\rangle(T,T_{rot},T_{v})$ is:
\begin{equation}
    \langle \hat{\epsilon}^{d}_v\rangle(T,T_{rot},T_{v}) =  \frac{\hat{\Phi}(\epsilon_d,0,1)+\hat{\Phi}(\epsilon_d^{\mx},\epsilon_d,2)}{\hat{\Hn}(\epsilon_d,0,1)+\hat{\Hn}(\epsilon_d^{\mx},\epsilon_d,2)}
    \label{avg_edv_QSS_CF2}
\end{equation}
where
\begin{equation}
\begin{split}
    \hat{\Phi}(\epsilon_i,\epsilon_j,n) = \hspace{3.0in} \\
    \cfrac{\exp[(-1)^{n-1}\delta]}{\hat{\Zn}(T_{rot},T_{v})} \frac{\exp\left[ \epsilon_i \hat{\zeta}_{rot}\right]  \hat{\gn}' (\hat{\zeta}_{vr}) - \exp\left[ \epsilon_j \hat{\zeta}_{rot}\right] \hat{\gn}' (\zeta_{v}-\epsilon_j \hat{\zeta}_{rot}/\epsilon_d )}{k_B \theta_{rot} \hat{\zeta}_{rot}} ;
    \label{Hnb_QSS_11}
\end{split}
\end{equation}
\begin{equation}
\begin{split}
    \hat{\Hn}(\epsilon_i,\epsilon_j,n) =  \hspace{3.0in} \\  \cfrac{\exp[(-1)^{n-1}\delta]}{\hat{\Zn}(T_{rot},T_{v})}  \frac{\exp\left[ \epsilon_i \hat{\zeta}_{rot}\right]  \hat{\gn} (\hat{\zeta}_{vr}) - \exp\left[ \epsilon_j \hat{\zeta}_{rot}\right] \hat{\gn} (\zeta_{v}-\epsilon_j \zeta_{rot}/\epsilon_d )}{k_B \theta_{rot} \hat{\zeta}_{rot}} ;
    \label{Ztvtr_QSS_12}
\end{split}
\end{equation}
and where the derivatives of $\hat{\gn}$ can be expressed in the following manner:
\begin{equation}
   \hat{\gn}'(x)= \sum_m \hat{\gn}'_m(x)
   \label{hatgderivativesum}
\end{equation}
\begin{equation}
    \hat{\gn}'_m(x)= \frac{\partial \hat{\gn}_m}{\partial x} = \hat{\gn}_m\frac{\partial \log \hat{\gn}_m}{\partial x} 
    \label{hatgderivative_m}
\end{equation}

\begin{equation}
\begin{split}
   \hat{\gn}'_m(x)=\hat{\gn}_m(x)\left\{ E_{m^-} - \frac{(m^+-m^-)  \theta_v^m\ k_B \exp[ (m^+-m^-) ( x \theta_v^m\ k_B +\hat{\delta}_v )]}{1-\exp[ (m^+-m^-)( x \theta_v^m\ k_B +\hat{\delta}_v)]} 
   + \frac{ \theta_v^m\ k_B \exp[x\  \theta_v^m\ k_B +\hat{\delta}_v ]}{1-\exp[x\  \theta_v^m\ k_B +\hat{\delta}_v  ]} \right \} 
\end{split}
\label{hatgderivative_CF2}
\end{equation}

As discussed in the article, we propose that the following simple approximation is accurate for the average rotational energy of dissociated molecules:
\begin{equation}
   \langle \epsilon^{d}_{rot} \rangle ^{NB}(T,T_{rot},T_{v}) =  \epsilon_d -  \langle \epsilon^{d}_{v} \rangle ^{NB}(T,T_{rot},T_{v})
    \label{avg_edrot_gen}
\end{equation}
The above proposition is also based on the finding in Ref.\cite{bender2015improved}, where the average internal energy of dissociating molecules is
approximately $\epsilon_d$ for the considered range of conditions.

\subsection{Calculation of the parameter $\Lambda$}
As seen in Eq.~B1, parameter $\Lambda$ controls the relative importance from both the overpopulation phase and the depleted QSS phase. As derived in the appendix (see Eqs. D1-D4 in the Sec.~D of Ref.~\cite{SinghCFDI}) is given by:
\begin{equation}
    \Lambda = \cfrac{\langle \epsilon_v \rangle - \langle \tilde{\epsilon_v} \rangle(T_v,T_0,T_{rot})}{ \langle \hat{\epsilon}_v \rangle (T_{rot},T_{v}) -\langle \epsilon_v\rangle }
\end{equation}

The parameter $\Lambda$ requires two quantities $\langle \tilde{\epsilon_v} \rangle$ and $ \langle \hat{\epsilon}_v \rangle$, which are mathematically described as:

\begin{equation}
\begin{split}
    \langle \tilde{\epsilon_v} \rangle (T_v,T_0,T_{rot}) = \cfrac{1}{\tilde{Z}(T_v,T_0,T_{rot})}\cfrac{T_{rot}}{\left[\theta_{rot} -\cfrac{ T_{rot}}{\theta_{rot}k_B}\hat{\delta}_{rot} \right]} \left \{ -\tilde{\gn} '\left(+\frac{1}{ k_B T_0}\right) \right. \\ \left. + \exp\left[-\cfrac{\epsilon_d^{\mx}}{k_B T_{rot}} +\epsilon_d^{\mx} \cfrac{\hat{\delta}_{rot}}{k_B \theta_{rot}} \right]  \times \tilde{\gn}' \left(+\frac{1}{ k_B T_0}+ \cfrac{1}{k_B T_{rot}} -\cfrac{\hat{\delta}_{rot}}{k_B \theta_{rot}}\right)\right \} 
    \end{split}
\end{equation}

\begin{equation}
\begin{split}
    \langle \hat{\epsilon_v} \rangle (T_{rot},T_{v}) = \cfrac{1}{\hat{\Zn}(T,T)}\cfrac{T_{rot}}{\left[ \theta_{rot} -\cfrac{ T_{rot}}{ \theta_{rot}k_B}\hat{\delta}_{rot} \right] }
    \left \{- \hat{\gn}' \left(-\frac{1}{ k_B T_v}\right)  + \exp\left[-\cfrac{\epsilon_d^{\mx}}{k_B T_{rot}} +\epsilon_d^{\mx} \cfrac{\hat{\delta}_{rot}}{k_B \theta_{rot}} \right] \right. \\ \left. \times \hat{\gn}' \left(-\frac{1}{ k_B T_v}+ \cfrac{1}{k_B T_{rot}} -\cfrac{\hat{\delta}_{rot}}{k_B \theta_{rot}}\right)\right \} 
    \end{split}
\end{equation}


\section{Simplifying the average energy of dissociating molecules, $\langle \epsilon_v^d \rangle$}{\label{simplify_edv_append} expression}
The expression for $\langle \epsilon_v^d \rangle$ is derived in Eq.~C1 in Sec.C of the appendix in Ref.~\cite{SinghCFDI}, which is: 
\begin{equation}
    \langle \hat{\epsilon}^{d}_v\rangle(T,T_{rot},T_{v}) =  \frac{\hat{\Phi}(\epsilon_d,0,1)+\hat{\Phi}(\epsilon_d^{\mx},\epsilon_d,2)}{\hat{\Hn}(\epsilon_d,0,1)+\hat{\Hn}(\epsilon_d^{\mx},\epsilon_d,2)}
    \label{avg_edv_QSS_1}
\end{equation}
where $\Phi(\epsilon_d,0,1)$ and $\Hn(\epsilon_d,0,1)$ are the contributions from  bound molecules, and $\Phi(\epsilon_d^{\mx},\epsilon_d,2)$ and $\Hn(\epsilon_d^{\mx},\epsilon_d,2)$ are the contributions from quasi-bound molecules. For simplification, recall assumption (iii) used in Sec.~\ref{der_simplify} of the article, which results in the following simplification.
\begin{equation}
\begin{split}
    \langle \hat{\epsilon}^{d}_v\rangle(T,T_{rot},T_{v}) =  \frac{C' \hat{\Phi}(\epsilon_d^{\mx},\epsilon_d,2)+\hat{\Phi}(\epsilon_d^{\mx},\epsilon_d,2)}{C' \hat{\Hn}(\epsilon_d^{\mx},\epsilon_d,2)+\hat{\Hn}(\epsilon_d^{\mx},\epsilon_d,2)}
    \\
    =  \frac{ \hat{\Phi}(\epsilon_d^{\mx},\epsilon_d,2)}{\hat{\Hn}(\epsilon_d^{\mx},\epsilon_d,2)}
    \end{split}
    \label{avg_edv_QSS_2}
\end{equation}
where using the expressions for $ \Phi(\epsilon_d^{\mx},\epsilon_d,2)$ and $\Hn(\epsilon_d^{\mx},\epsilon_d,2)$ given in Eqs.~\ref{Hnb_QSS_11} and \ref{Ztvtr_QSS_12} respectively, we can obtain:
\begin{equation}
\begin{split}
    \langle \hat{\epsilon}^{d}_v\rangle(T,T_{rot},T_{v})    =  \cfrac{\hat{\gn}' (\hat{\zeta}_{vr}) }{\hat{\gn} (\hat{\zeta}_{vr})}
    \end{split}
    \label{avg_edv_QSS_2}
\end{equation}
The expression for $\hat{\gn}' (\hat{\zeta}_{vr})$ can be obtained from the expression of $\hat{\gn}' (x)$ in Eq.~\ref{hatgderivative_CF2} by substituting $x = \hat{\zeta}_{vr}$
\begin{equation}
\begin{split}
   \hat{\gn}'_m(x)=\hat{\gn}_m(x)\left\{ E_{m^-} - \frac{(m^+-m^-)  \theta_v^m\ k_B \exp[ (m^+-m^-) ( x \theta_v^m\ k_B +\hat{\delta}_v )]}{1-\exp[(m^+-m^-)( x \theta_v^m\ k_B +\hat{\delta}_v)]} 
   + \frac{ \theta_v^m\ k_B \exp[x\  \theta_v^m\ k_B +\hat{\delta}_v ]}{1-\exp[x\  \theta_v^m\ k_B +\hat{\delta}_v  ]} \right \} 
\end{split}
\label{hatgderivative}
\end{equation}
Using the same SHO (assumption vi in Sec.~\ref{der_simplify}), we can use an effective $\theta_v^{\ef}$ and simplify the expression in $\langle \hat{\epsilon}^{d}_v\rangle(T,T_{rot},T_{v})$ as:
\begin{equation}
\begin{split}
    \langle \hat{\epsilon}^{d}_v\rangle(T,T_{rot},T_{v})    =  E_{0} - \frac{\epsilon_d \exp\left[  \epsilon_d \hat{\zeta}_{vr}  + \cfrac{\hat{\delta}_v \epsilon_d}{k_B \theta_v^{\ef}}\right]}{1-\exp\left[\epsilon_d \hat{\zeta}_{vr}  + \cfrac{\hat{\delta}_v \epsilon_d}{k_B \theta_v^{\ef}}\right]} 
   + \frac{ \theta_v^{\ef}\ k_B \exp\left[ \hat{\zeta}_{vr} \  \theta_v^{\ef}\ k_B +\hat{\delta}_v \right]}{1-\exp\left[ \hat{\zeta}_{vr}\  \theta_v^{\ef}\ k_B +\hat{\delta}_v  \right]}
   \\
   =
   E_{0} - \frac{\epsilon_d }{\exp\left[-\epsilon_d \hat{\zeta}_{vr}  - \cfrac{\hat{\delta}_v \epsilon_d}{k_B \theta_v^{\ef}}\right]-1} 
   + \frac{ \theta_v^{\ef}\ k_B }{\exp\left[ -\hat{\zeta}_{vr}\  \theta_v^{\ef}\ k_B -\hat{\delta}_v  \right]-1}
    \end{split}
    \label{avg_edv_QSS_3}
\end{equation}

which can be expressed using $T_F'$ as
\begin{equation}
\begin{split}
    \langle \hat{\epsilon}^{d}_v\rangle(T,T_{rot},T_{v})    =   E_{0} - \frac{\epsilon_d }{\exp\left[\cfrac{\epsilon_d}{k_B T_F'} \right]-1} 
   + \frac{ \theta_v^{\ef}\ k_B }{\exp\left[ \cfrac{\theta_v^{\ef}}{T_F'}  \right]-1}
    \end{split}
    \label{avg_edv_QSS_4}
\end{equation}
\end{widetext}

\subsection{New Non-equilibrium Model Parameters}
All of the parameters needed in the model equations are shown in Table I. 
\begin{table}[H]
\centering 
\caption{\label{tab:table-name} Constants for approximation of \textit{ab initio} energies and parameters required in the model.}
\begin{tabular}{ | m{13em} | m{4.2cm}| }
\hline
Vibrational energy  & $\theta_v^{I}=3390$ K for $\vv \in [0,9) $,\\ 
(SHO) &  $\theta_v^{II}=0.75\  \theta_v^{I}$  for $\vv \in [9,31) $,  \\ 
 &  $\theta_v^{III}=0.45\  \theta_v^{I}$  for $\vv \in [31,55) $  \\ 
\hline
Rotational energy (Rigid Rotor) & $\theta_{rot} = 2.3$  K \\ 
\hline
Centrifugal barrier & $\theta_{CB} = 0.27$  \\ 
\hline
Diatomic energies & $\epsilon_d = 9.91 \text{eV} , \epsilon_d^{\max}=14.5 \text{eV}$  \\ 
\hline
Reaction probability & $C_{1} = 8.67\times 10^{-5}, \ \alpha =1.04 $, 
\\
                      &  $ \beta=5.91, \ \gamma=3.49,\ \delta = 1.20$  
\\ 
\hline
Non-Boltzmann distributions \cite{Singhpnas} & $\lambda_{1,v}=0.080, \lambda_{1,j}=4.33\times 10^{-5} $  \\ 
\hline
\end{tabular}
\label{parameters}
\end{table}
\bibliography{Bib_CFD_DSMC}
\end{document}